\documentclass[preprint, twocolumn]{aastex701}

\usepackage{amsmath}
\usepackage[T1]{fontenc}

\newcommand{\cuharm}{\texttt{cuHARM}}

\newcommand{\Ihat}{\hat{I}}
\newcommand{\nhat}{\hat n}
\newcommand{\jhat}{\hat{j}}
\newcommand{\alphahat}{\hat{\alpha}}
\newcommand{\zetahat}{\hat{\zeta}}
\newcommand{\psihat}{\hat{\psi}}

\newcommand{\Ibar}{\bar{I}}

\newcommand{\Ebar}{\bar{E}}
\newcommand{\nbar}{\bar{n}}
\newcommand{\jbar}{\bar{j}}
\newcommand{\alphabar}{\bar{\alpha}}

\newcommand{\g}{\mathrm{g}}

\newcommand{\K}{\mathrm{K}}
\newcommand{\s}{\mathrm{s}}

\newcommand{\arad}{a_\mathrm{rad}}
\newcommand{\mprot}{m_\mathrm{p}}
\newcommand{\kB}{k_\mathrm{B}}
\newcommand{\Msun}{M_\odot}

\newcommand\diff{\mathop{}\!\mathrm{d}}

\newcounter{tr}
\ifnum \value{tr}>5

\newcommand{\deletedD}[1]{{\color{red} Damien - Deleted: } \sout{#1}}

\newcommand{\authorcommentD}[1]{{\color{purple} Damien - Comment :} {\color{cyan} #1}}

\else

\newcommand{\deletedD}[1]{}

\newcommand{\authorcommentD}[1]{}

\fi

\graphicspath{{./}{figures/}}

\begin{document}

\title{Direct Solution of the Time-Dependent Covariant Radiative Transfer Equation and its Coupling to General Relativistic Magnetohydrodynamics with \cuharm{}}


\author[orcid=0000-0001-6180-801X,gname=John,sname=Wallace]{John Wallace}
\affiliation{Bar-Ilan University, Ramat Gan, Israel}
\email[show]{john.wallace@biu.ac.il}

\author[orcid=0000-0003-4477-1846,gname=Damien,sname=Begue]{Damien B\'{e}gu\'{e}}
\affiliation{Bar-Ilan University, Ramat Gan, Israel}
\email[show]{begueda@biu.ac.il}

\author[orcid=0000-0001-8667-0889,gname=Asaf,sname=Peer]{Asaf Pe'er}
\affiliation{Bar-Ilan University, Ramat Gan, Israel}
\email[show]{asaf.peer@biu.ac.il}

\begin{abstract}

In this paper we present a major update to the general relativistic magnetohydrodynamics (GRMHD) code
\cuharm, which adds fully covariant treatment of radiation transport and the subsequent radiation backreaction
on the dynamics of the fluid. For the radiative calculations, we discretize and solve the radiation transfer equation on a geodesic grid, in order to resolve the angular distribution of the radiation field everywhere in space. This allows for detailed treatment of non-isotropic radiation fields, which is crucial for accurately
resolving regions of intermediate optical depth. We present the equations solved, the numerical methods used, and standard tests used to verify the
different aspects of a radiation hydrodynamics code, in particular radiation transport and radiation-fluid
interaction. We present an application of the code to the case
of black hole radiative accretion. This new radiation module is fully GPU-accelerated and represents a major advance
in the capabilities of \cuharm{}. 

\end{abstract}


\section{Introduction \label{sec:intro}}

General relativistic magnetohydrodynamics (GRMHD) simulations are well-established in the study of accreting
black hole systems, where they have been used extensively to study the turbulent dynamics found in these scenarios
\citep[see e.g.][]{GMT03, AFS05, NGM06, FGM12, EPH15, WSG16, POM17, LCI22, BPZ23}. These simulations 
are computationally intensive, which limits not just the resolution of the simulations, but also the physical processes which can be feasibly included therein. Recent improvements in the availability of computational power have made it possible to include the effects of radiative transfer on the dynamics of the accreting plasma, which would represent a major step forward for the accuracy of GRMHD modeling of accreting black hole systems \citep{MTS14, ATO20, WMJ23}. The inclusion of radiative processes in GRMHD codes opens the door to a far wider parameter space for study; in particular, for situations in which the influence of radiation has a significant impact on the dynamics and evolution of the accretion disk. Such scenarios are likely to arise in ``slim disks'' \citep{ACL88} for example, where the accretion rate is a large enough fraction (say, greater than a few per cent) of the Eddington accretion rate, which results in regions of high density (and therefore high optical depth) close to the central black hole, leading to potentially significant interaction between plasma and radiation.

Radiation MHD has long been recognized as a crucial aspect of accurately modeling accretion flows.
It is though, very computationally expensive. This problem is exacerbated by the need to set the unit scales
of the simulation from the outset, due to the dependence of
certain radiative quantities, notably optical depth and emissivity, on temperature and density in physical
units. In particular, this requires the central black hole (BH) mass to be specified at the outset, which is not the case for non-radiative simulations. As such, while GRMHD simulations can be run in a scale-free setup, greatly narrowing the number of
simulations required to study different systems, the addition of radiation removes this freedom. Where
a single GRMHD simulation of a particular configuration could be scaled with BH mass, from
stellar mass, through intermediate mass to supermassive BHs, the addition of radiation necessitates
a new simulation for each different BH mass, even when all other parameters (e.g. spin,
magnetic field configuration) are identical.

A number of approaches have been proposed for the direct (as opposed to post-processed) treatment of radiation in GRMHD simulations. 
The most commonly used approaches today are moment-based methods, most notably the M1 closure method \citep[][for a very partial list]{Levermore1984, Mihalas2, Shibata+11, SNT13, MTS14, Fragile+14, Takahashi+16, LMT22, Cheong+23}. 
In this method, the radiative transfer equation is not solved directly, but rather its first and second moments. Since the $n^{th}$ moment of the radiation stress-energy tensor depends only on the $(n + 1)^{th}$ moment, 
it is enough to specify the form of the third moment ($P^{ij}$) in order to close the system of equations. This is most commonly done using the M1 prescription introduced in \citet{Levermore1984}. This method is proven to converge to the correct result in the limits of high ($\tau \gg 1$) and low ($\tau \ll 1$) optical depths, although it cannot accurately describe non-isotropic radiation fields, which occur at intermediate optical depths.  

Another commonly used method is the Monte Carlo method, which attempts to generate a statistically accurate sample of the radiation field \citep{RDG15,RD20, Foucart+21, Roth+22, Kawaguchi+23}. A drawback of this method is that it can become extremely expensive in regions of high optical depth, where multiple scattering takes place. An example of a hybrid scheme was introduced by \citet{Foucart18}, where Monte Carlo techniques are used to
sample the local radiation field, to provide better closures for a moment method. 

A third method is the grid-based Boltzmann approach \citep{Nagakura+17, ATO20, WMJ23}. In this method, the radiation field
is evolved directly on a discrete angular grid. This effectively results in a 6-dimensional system to be solved for the MHD
primitives and the angular distribution of the radiation. A variant is the Variable Eddington Tensor
(VET) method, which uses an angular description of the radiation field to calculate the Eddington tensor, which then provides the closure to be used in the moment method \citep{SMN92, JSD12, ATO20, MFK22, MPJ25}.  
 
There are several advantages of a grid-based Boltzmann solver over other methods. Moment methods such as M1 fail
to adequately capture directional effects of radiation, and therefore lose accuracy in regions where the radiation
field is anisotropic \citep{Mihalas2}. In contrast, for a grid-based solution, anisotropies in the
radiation field are handled naturally, as long as sufficient angular resolution is available \citep{ATO20}.
This is crucial when studying systems in regions of intermediate optical depths, where the radiation field
is highly anisotropic \citep[e.g.][]{Beloborodov11}, as well as when angle-dependent cross sections
becomes of importance. 

While Monte Carlo methods can offer good sampling of the angular distribution of radiation at low optical depths, they suffer greatly from lack of efficiency at higher optical depths due to the large number of samples required to accurately model the statistical distribution of the radiation field \citep{RDG15,RD20}. Grid-based approaches, on the other hand, can be used flexibly across optical depth ranges. 

Although such deterministic grid-based approaches have existed for some time \citep{JSD14, OT16, Jia21}, their application to fully general-relativistic regimes is a more recent development, enabled by increased availability of computational power and in particular GPU-acceleration \citep{Shibata2014, DG20, ATO20, WMJ23}. Given its advantages over moment-based methods, we chose to adopt the Boltzmann framework within our GR-MHD code \cuharm\ \citep{BPZ23}.

The computational cost of such a scheme is considerable. 
However, the GPU-based approach of \cuharm\ lends itself well to overcoming this obstacle. Each relevant kernel of \cuharm\ is highly optimized to minimize computational time, making it one of the most efficient codes of its kind.  Thus, our GPU-accelerated code puts the required number of simulations to carry out a comprehensive study of parameter space within reach over the course of a reasonable time scale.

In systems accreting at high accretion rates, close to or even above the Eddington limit, the effects of radiative feedback can have important consequences for the dynamics of the accretion flow. Radiative cooling can have a significant effect on the disk structure at much lower accretion rates, lower than $10^{-5}$ Eddington \citep{Liska+22b, SBP25}. High accretion rates can be reached in ultraluminous X-ray sources (ULXs) \citep{Watarai2001}, as well as in radio-quiet active galactic nuclei (AGNs) \citep{Sikora2007} and X-ray binaries \citep[][and references therein]{AF13}. Radiative effects can also play an important role in super-Eddington transients such as tidal disruption events (TDEs) and gamma-ray bursts (GRBs), where the extreme luminosities mean that radiation pressure is the dominant factor driving the dynamics \citep{Pir99, Peer15, KZ15}. 


This paper is organized as follows: in Section~\ref{sec:method}, we introduce the covariant radiative transfer equation, along with the necessary frames and coordinate transforms required to solve the equations of general relativistic radiation magnetohydrodynamics. In Section~\ref{sec:numerical_scheme}, we outline our discretization of the radiative transfer equation, the structure of our computational grid, and the numerical fluxes used to evolve the specific intensity $I_\nu$. In addition, we describe our method of accounting for radiation-matter interaction. In Section~\ref{sec:tests}, we present several tests of our code, demonstrating that we retrieve the correct behavior for both the radiation transport sector as well as the interaction between radiation and matter. In Section~\ref{sec:results}, we present the results of a complete radiative GRMHD simulation of an accreting torus, demonstrating the capability of \cuharm{} to handle large scale radiative simulations. In Section~\ref{sec:before_conclusion}, we discuss some limitations of our approach and how GPUs can handle the heavy workload of simulating radiative fluids with full angular and spectral information for the radiation field.  Finally, in Section~\ref{sec:discussion} we summarize our work and discuss possible future development directions.

\section{General Relativistic Radiation Magnetohydrodynamics}
\label{sec:method}

We introduce here the set of equations solved. We follow closely the prescription introduced by \citet{DG20, WMJ23}.
When appearing, $\nu$ refers to the frequency
of the specific intensity, $I_\nu$. 

Throughout the paper, Einstein summation convention is used. Greek indices, apart from $\nu$, which represents the radiation frequency, run from 0 to 3 irrespective of their
decorator. Latin indices, apart from $l$ and $s$, which we reserve for describing the numerical discretizations (see Section~\ref{sec:numerical_scheme}), run from 1 to 3. In addition, we work in units where $G = c = 1$.

\subsection{Coordinate Frames and Transformations \label{sec:coordinate_transforms}}

Our method of calculation of radiative transfer and radiation-matter interaction requires the use of three coordinate systems: the coordinate frame, the tetrad frame and the comoving fluid frame. In this section, we describe the latter two coordinate systems and the transformations between them. Details of the coordinate frames used are provided for the individual test cases presented below. 

\subsubsection{Tetrads \label{sec:tetrads}}

Radiative transport calculations are significantly simplified by making use of a tetrad frame, which is an orthonormal frame satisfying
\begin{equation}
    e^{\gamma}_{\hat{\alpha}} e^{\delta}_{\hat{\beta}} g_{\gamma \delta} = \eta_{\hat{\alpha} \hat{\beta}}.
    \label{eq:tetrad_def}
\end{equation}
Namely, tetrads represent a transformation to a frame which is locally Minkowski. Here, $e^{\delta}_{\hat{\alpha}}$ is a set of 4 basis vectors  (one for each $\hat \alpha \in {0\ldots3}$) that, at any location, relate the coordinate frame to the tetrad frame. It is important to note that the choice of a tetrad is generally not unique for a given metric or coordinate system, and can often be chosen in such a way as to simplify calculations involved in the problem at hand. Throughout this work we use a ``hat'' on both indices
(i.e. $\hat{\alpha}$, $\hat{\beta}$, etc.) and quantities (i.e. $\hat I_{\hat \nu}$) to refer to quantities in the tetrad
frame, while non-decorated indices represent coordinate-frame quantities. The corresponding inverse transformations will also
be useful throughout our work and can similarly be calculated using
\begin{equation}
    g_{\gamma \delta} = e^{\hat{\alpha}}_{\gamma} e^{\hat{\beta}}_{\delta} \eta_{\hat{\alpha} \hat{\beta}}.
    \label{eq:inverse_tetrad_def}
\end{equation}
Here, $e^{\hat{\alpha}}_{\epsilon}$ are the components of the tetrad vector in the coordinate frame, satisfying $e^{\hat{\alpha}}_{\epsilon}  e^{\gamma}_{\hat{\beta}} = \delta_{\hat \beta}^{\hat \alpha} \delta_\epsilon^\gamma$. 
Quantities in the coordinate basis can be transformed to and from the tetrad frame using the relations:
\begin{align} \left \{
    \begin{aligned}
        \hat{x}^{\hat{\alpha}} &= e^{\hat{\alpha}}_{\delta} x^{\delta} \\
        x^{\delta} &= e^{\delta}_{\hat{\alpha}} \hat{x}^{\hat{\alpha}}
    \end{aligned} \right.
\label{eq:tetrad_transform}
\end{align}
respectively.

We further require that the time-like direction in the tetrad frame ($\hat{x}^{\hat 0}$) be orthogonal to surfaces of constant $x^0$ (coordinate-time), and that it is future-pointing. Mathematically, this can be expressed as:
\begin{equation}
    e^{\alpha}_{\hat 0} = -\frac{1}{\sqrt{-g^{00}}} g^{0 \alpha}.
    \label{eq:tetrad_restriction}
\end{equation}

The choice of tetrad is crucial in this numerical method, and there often exists a tetrad which can minimize the numerical diffusion or exploit symmetries of the problem in question to increase the accuracy of the numerical scheme.  For instance, for a flat spacetime, the natural choice of tetrad in Cartesian coordinates is simply the identity matrix (see Section~\ref{sec:tetrad_choice_test} for a comparison of other possible tetrad choices). We describe in Appendix~\ref{app:coord} the tetrad choices used in our work. It is important to note that any choice of a set $e^{\delta}_{\hat{\alpha}}$ that satisfies the relation in Equation~\eqref{eq:tetrad_def} is a valid choice of tetrad and can be used to carry out the necessary computations. Similar to \citet{DG20}, in this work we only consider a stationary metric, which leads to stationary tetrads (i.e. $\partial_0 e^{\delta}_{\hat \alpha} = 0$). This simplifies certain aspects of the implementation and places further constraints on our choice of tetrad.

\subsubsection{The Comoving Frame \label{sec:comoving_frame}}

Many of the calculations involved in the interaction are carried out in the fluid rest-frame, in which the plasma is isotropic
by definition. Quantities in this frame will be decorated with a bar (e.g. $\Ibar_{\bar \nu}$), and indices are also barred (e.g. $\alphabar$, $\bar{\beta}$, etc.).
The transformation between the tetrad frame and the fluid frame
is given by a Lorentz transformation, written as:

\begin{equation}
    \Lambda^{\bar{\delta}}_{\hat{\gamma}} =
    \begin{pmatrix}
        \hat{u}^{\hat{0}} & -\hat{u}^{\hat{1}} & -\hat{u}^{\hat{2}} & -\hat{u}^{\hat{3}} \\
        \hat{u}_{\hat{1}} & \frac{\hat{u}^{\hat{1}} \hat{u}_{\hat{1}}}{1 + \hat{u}^{\hat{0}}} + 1 & \frac{\hat{u}^{\hat{2}} \hat{u}_{\hat{1}}}{1 + \hat{u}^{\hat{0}}} & \frac{\hat{u}^{\hat{3}} \hat{u}_{\hat{1}}}{1 + \hat{u}^{\hat{0}}} \\
        \hat{u}_{\hat{2}} & \frac{\hat{u}^{\hat{1}} \hat{u}_{\hat{2}}}{1 + \hat{u}^{\hat{0}}} & \frac{\hat{u}^{\hat{2}} \hat{u}_{\hat{2}}}{1 + \hat{u}^{\hat{0}}} + 1 & \frac{\hat{u}^{\hat{3}} \hat{u}_{\hat{2}}}{1 + \hat{u}^{\hat{0}}} \\
        \hat{u}_{\hat{3}} & \frac{{\hat u}^{\hat{1}} {\hat u}_{\hat{3}}}{1 + {\hat u}^{\hat{0}}} & \frac{{\hat u}^{\hat{2}} {\hat u}_{\hat{3}}}{1 + {\hat u}^{\hat{0}}} & \frac{{\hat u}^{\hat{3}} {\hat u}_{\hat{3}}}{1 + {\hat u}^{\hat{0}}} + 1
    \end{pmatrix},
    \label{eq:fluid_to_coordinate}
\end{equation}
where ${\hat u}^{\hat \delta} = e^{\hat \delta}_{\gamma} u^\gamma$ is the 4-velocity of the flow expressed in the tetrad frame, calculated using Equation~\eqref{eq:tetrad_transform}.


\subsection{Covariant Radiative Transfer Equation}

Consider the future-oriented null vector of norm unity aligned with the photon momentum in the tetrad frame $\hat n^{\hat \delta}$. Its components are:
\begin{equation}
    \hat n^{\hat \delta} = ( 1, \sin(\hat \zeta ) \cos(\hat \psi), \sin(\hat \zeta ) \sin(\hat \psi), \cos (\hat \zeta)),
    \label{eq:n_vector_definition}
\end{equation}
where $\hat \zeta$ and $\hat \psi$ are polar angles representing the local direction of the photon in the tetrad frame. Although
$\hat n^{\hat \delta}$ does not depend on spatial position, the corresponding coordinate frame {$n^\alpha = e_{\hat \delta} ^\alpha \hat n ^{\hat \delta}$} is position-dependent through the transformation given by Equation~\eqref{eq:tetrad_transform}.

We further define the symbols:
\begin{subequations}
    \begin{align}
        r^{\hat \nu} &= -{\hat \nu} \nhat^{\hat \gamma} \nhat^{\hat \delta} \omega^{\hat 0}_{\hat \gamma \hat \delta} \label{eq:n_nu_equation}, \\
        r^{\hat \zeta} &= -\frac{1}{\sin{\zetahat}} \nhat^{\hat \gamma} \nhat^{\hat \delta} \left(\nhat^{\hat 0} \omega^{\hat 3}_{\hat \gamma \hat \delta} - \hat{n}^{\hat 3}\omega^{\hat 0}_{\hat \gamma \hat \delta} \right) \label{eq:n_zeta_equation}, \\
        r^{\hat \psi} &= \frac{1}{\sin^2{\zetahat}} \nhat^{\hat \gamma}  \nhat^{\hat \delta} \left(\nhat^{\hat 2}\omega^{\hat 1}_{\hat \gamma \hat \delta} - \nhat^{\hat 1}\omega^{\hat 2}_{\hat \gamma \hat \delta} \right) \label{eq:n_psi_equation},
    \end{align}
    \label{eq:angular_n_vectors}
\end{subequations}
where the Ricci rotation coefficients, $\omega^{\hat{\alpha}}_{\hat \gamma \hat \delta}$, are given by:
\begin{equation}
    \omega^{\hat{\alpha}}_{\hat \gamma \hat \delta} = e^{\hat \alpha}_\beta e^\mu_{\hat \gamma} \nabla_\mu e^\beta_{\hat \delta},
    \label{eq:ricci_rotation_coefficients}
\end{equation}
where $\nabla_\mu$ is the covariant derivative. These coefficients play the role of the connection in the tetrad basis, and hence encode the rotation of the tetrad frame with spatial position,
which allows us to track the trajectory of a light ray as it propagates through an arbitrary spacetime.
We note that the covariant derivatives are calculated numerically using the finite difference method for the derivative and accounting for the contributions of the connection terms. 

Radiation transfer in general curvilinear coordinates is described by the covariant radiative
transfer equation (hereinafter RTE). With the above quantities defined, following \citet{DG20}, we can write the RTE in flux-conservative form as:
\begin{align}
    &\nabla_{\alpha} \left( n^{\alpha} n_{\beta} \Ihat_{\hat \nu} \right) + \partial_{\hat \nu} \left( r^{\hat \nu} n_{\beta} \Ihat_{\hat \nu} \right)  \nonumber \\ 
    +&\frac{1}{\sin{\zetahat}}\partial_{\zetahat} \left( \sin{\zetahat} r^{\zetahat} n_{\beta} \Ihat_{\hat \nu} \right) + \partial_{\psihat} \left( r^{\psihat} n_{\beta} \Ihat_{\hat \nu} \right) \label{eq:covariant_radiative_transfer} \\
    =&  n_{\beta} \left( \jhat_{\hat \nu}\ - \alphahat_{\hat \nu} \Ihat_{\hat \nu} \right), \nonumber 
\end{align}
where $\Ihat_{\hat \nu}$ is the specific intensity of the radiation in the tetrad frame, and $n^{\alpha}$ and $n_{\alpha}$ are
respectively the contravariant and covariant vectors along the direction of propagation of radiation in the coordinate frame\footnote{Note that this form of Equation~\eqref{eq:covariant_radiative_transfer} holds in any arbitrary frame.}.  
On the left hand side of this equation, the second term describes how
radiation frequency changes as it propagates in a curved space-time, while the third and fourth terms account for the change of direction
of photon propagation in the polar and azimuthal angles, respectively. On the right hand side, $\hat j_{ \hat \nu}$
is the emissivity and ${\hat \alpha}_{\hat \nu}$ is the absorptivity, as measured in the local tetrad frame \citep[see e.g.][]{Mihalas2}. It is important to note that the quantities ${\hat j}_{\hat \nu}$ and ${\hat \alpha}_{\hat \nu}$, necessary to solve Equation~\eqref{eq:covariant_radiative_transfer}, require physical units to evaluate. Unit conversions are provided in Section~\ref{sec:mass_scaling}. Finally, $\beta$ can be chosen arbitrarily, and corresponds to the conservation of different quantities. We choose $\beta = 0$ to conserve a form of energy in the radiation transfer \citep[see][for further details]{DG20}.

In this paper, we limit the analysis to the gray (i.e. frequency integrated) RTE. We introduce the gray intensity $\hat I$:
\begin{align}
    {\hat I} &= \int_0^\infty {\hat I}_{\hat \nu} \diff {\hat \nu}  \label{eq:gray_intensity},
\end{align}
alongside the frequency-integrated emission and absorption coefficients

\begin{align}
    {\hat j} &= \int_0^\infty {\hat j}_{\hat \nu} \diff {\hat \nu},  & ~~~     {\hat \alpha} &= \frac{1}{{\hat I}}\int_0^\infty {\hat \alpha}_{\hat \nu} \diff {\hat \nu}. \label{eq:gray_absorptivity}
\end{align}

\subsection{The MHD Sector}
In order to incorporate the effects of radiation into our GRMHD calculations, we modify the equations of ideal GRMHD to include the contribution of the radiation onto the dynamics via the radiation
four-force density $G^\alpha$, which represents the exchange of energy-momentum between the radiation and the fluid.
The radiation 4-force is defined as:
\begin{align}
    G_\alpha = -\oint \diff {\hat \Omega} \int \diff {\hat \nu} ~~ n_\alpha ({\hat j}_{\hat \nu} - {\hat \alpha}_{\hat \nu} {\hat I}_{\hat \nu}).  
    \label{eq:galpha}
\end{align}
Since $\hat I_{\hat \nu}$ is evolved via Equation~\eqref{eq:covariant_radiative_transfer}, the radiation 4-force is readily calculated. Practically, while we could use the definition provided in Equation~\eqref{eq:galpha}, to achieve a fully conservative scheme (energy and momentum) we instead calculate the radiation 4-force using the temporal variation of the radiation stress energy tensor $R^{0 \alpha}$, calculated from $\hat I$ using 
\begin{align}
    R^{\alpha \beta} = \int n^{\alpha} n^{\beta} {\hat I} \diff {\hat \Omega}.
    \label{eq:R}
\end{align}
Description of the implementation appears in Section~\ref{sec:interaction} below.

The GRMHD sector of \cuharm\ solves the conservation of mass, energy and momentum equations, as well as Maxwell's equations,
\begin{subequations}
    \begin{align}
        \nabla_\alpha \left( \rho u^{\alpha} \right) &= 0, \label{eq:mass_conservation}\\
        \nabla_\alpha \left( T^{\alpha}_{~~\beta} \right) &= G_{\beta}, \label{eq:energy_momentum_conservation}\\
        \nabla_\alpha \left( \star F^{\alpha \beta} \right) &= 0. \label{eq:maxwell}
    \end{align}
\end{subequations}
Here, $\rho$ is the gas density, $u^\alpha$ is its 4-velocity, $T^{\alpha}_{~~\beta}$ is the matter stress energy tensor,
and $\star F^{\alpha \beta}$ is the dual of the Faraday tensor. The matter stress energy tensor is given by:
\begin{align}
T^{\alpha}_{~~\beta} = (\rho + u_g + p_g)u^\alpha u_\beta + \left ( p_g + \frac{b^2}{2}\right ) \delta ^\alpha_\beta - b^\alpha b_\beta,
\end{align}
where $u_g$ is the gas internal energy density, ${p_g = (\gamma-1)u_g}$ is the gas pressure assuming an ideal gas with an adiabatic index $\gamma$, and $b^2 = b^\alpha b_\alpha$. The magnetic field four-vector, $b^\alpha$, is defined from
the dual of the Faraday tensor $F^{\alpha \beta}$ as $b^\alpha =  \star F^{\alpha \beta} u_\beta$, which under the condition of
ideal MHD, $u_\alpha F^{\alpha \beta} = 0$, can be expressed as:
\begin{align}
    \star F^{\alpha \beta} = b^\alpha u^\beta - b^\beta u^\alpha.
\end{align}
Note that we absorb the factor $4\pi$ into the definition of the magnetic field.

Since we solve the RTE directly for the specific intensity $I_\nu$ from the emission and absorption processes, we use the variation of $I_\nu$ to calculate the variation of the radiation stress energy tensor $R^{0 \alpha}$ from which the radiation four-force 
$G^\alpha$ is directly derived (see below). Our approach is different from the frequently used technique based
on the method of radiation moments coupled to a closure relation \citep[see e.g.][]{Shibata+11, SNT13, MTS14, Fragile+14, Takahashi+16, LMT22, Cheong+23}. In that method, one must assume that the first two moments of the intensity provide an accurate description of the radiation field, and that the comoving radiation 4-force is a function of these two moments, thereby relying on angle-integrated emission and absorption processes. One then uses the evolutionary equation, $\nabla_{\alpha}R^{\alpha \beta} = - G^{\beta}$, to solve for the evolution of the radiation moments. Note that in this approach, the intensity is not calculated, only its first two moments. 

\section{Numerical Scheme}
\label{sec:numerical_scheme}

We describe here the numerical scheme employed to solve the radiative transfer equation using the finite volume method. From here on, we will refer to the local momentum grid as the ``angular grid'', or ``geodesic grid'' (see Section~\ref{sec:geodesic grid}).
The numerical techniques used to solve the MHD sector are fully described in \citet{BPZ23} and closely follow the implementation of the original work by \citet{GMT03}.

We will use the index $l$ to denote quantities expressed at the center of a spatial cell.
Consequently, when we use $l \pm 1/2$, it is understood that this quantity is expressed at the (spatial) cell boundary. Since for the radiative calculations we use an angular grid (see Section~\ref{sec:geodesic grid} below), the spatial indices are
always accompanied by an angular index, $s$, corresponding to a direction of propagation of radiation within the spatial cell $l$. We will also use the
index $s \pm 1/2$ to represent a quantity expressed at the boundary of the \textit{angular} cell. Note that with our choice of angular grid
(see below, Section~\ref{sec:geodesic grid}), the factor $1/2$ loses its positional meaning, and instead refers simply to the relevant boundary.

\subsection{Finite Volume Discretization}

In order to discretize Equation~\eqref{eq:covariant_radiative_transfer} following the finite volume methodology, we define a mesh in space and momentum (see Sections~\ref{sec:spatial_discretization} and \ref{sec:geodesic grid} below).
We then integrate Equation~\eqref{eq:covariant_radiative_transfer}, with $\beta=0$, over solid angle, frequency
and spacetime volume, to obtain: 
\begin{multline}
    \frac{1}{\Delta t} \left[ \langle {n}^0 {n}_0 \Ihat_{\hat{\nu}} \rangle_{V\hat{\Omega}} \right]_{x_0^-}^{x_0^+}
    + \sum_a \frac{1}{\Delta V} \left[ \Delta A_a \langle {n}^a {n}_0 \Ihat_{\hat{\nu}} \rangle_{t A_a \hat{\Omega}} \right]_{x_a^-}^{x_a^+}
    \\ + \frac{1}{\Delta \hat{\nu}} \left[ \langle r^{\hat{\nu}} {n}_0 \Ihat_{\hat{\nu}} \rangle_{t V\hat{\Omega}} \right]_{\hat{\nu}_0^-}^{\hat{\nu}_0^+} + \sum_{\hat{l}} \frac{1}{\Delta \hat{\Omega}} \left[ \Delta \lambda_{\hat{l}} \langle r^{\hat{l}} {n}_0 \Ihat_{\hat{\nu}} \rangle_{t V \lambda_{\hat{l}}} \right]_{\lambda_{\hat{l}}^-}^{\lambda_{\hat{l}}^+}
    \\ = \langle {n}_0 (\jhat_{\hat{\nu}} - \alphahat_{\hat{\nu}} \Ihat_{\hat{\nu}}) \rangle_{t V \hat{\Omega}},
    \label{eq:radiative_transfer_discrete_non_gray}
\end{multline}
where $a \in \{1,2,3\}$, ${\hat l} \in \{{\hat \zeta}, {\hat \psi}\}$  and
square brackets ``$\left[ x \right]$'' represent numerical fluxes, which are defined in Section~\ref{sec:fluxes}. The surface area of a spatial cell in the direction corresponding to $x^a$ is denoted by $\Delta A_a$, and $\Delta \lambda_{\hat l}$ is the a length of a geodesic cell boundary (see below).
Here, we have used the same notation as in \citet{WMJ23}, namely:
\begin{align*}
&\Delta t = \int \diff x^0, &&\langle \cdot \rangle_t = \frac{1}{\Delta t} \int \cdot \, \diff t, \\
&\Delta V = \int \sqrt{-g} \, \diff x^1 \diff x^2 \diff x^3, &&\langle \cdot \rangle_V = \frac{1}{\Delta V} \int \cdot \, \sqrt{-g} \, \diff x^1 \diff x^2 \diff x^3, \\
&\Delta A_a = \int \sqrt{-g} \, \diff x^b \diff x^c, &&\langle \cdot \rangle_{A_a} = \frac{1}{\Delta A_a} \int \cdot \, \sqrt{-g} \, \diff x^b \diff x^c, \\
&\Delta \hat{\nu} = \int \diff \hat{\nu}, &&\langle \cdot \rangle_{\hat{\nu}} = \frac{1}{\Delta \hat{\nu}} \int \cdot \, \diff \hat{\nu}, \\
&\Delta \hat{\Omega} = \int \sin{\zetahat} \, \diff \hat{\zeta} \, \diff \hat{\psi}, &&\langle \cdot \rangle_{\hat{\Omega}} = \frac{1}{\Delta \hat{\Omega}} \int \cdot \, \sin{\zetahat} \, \diff \hat{\zeta} \, \diff \hat{\psi}, \\
&\Delta \lambda_{\hat{l}} = \int \sin{\zetahat} \, \diff \lambda_{\hat{m}}, &&\langle \cdot \rangle_{\lambda_{\hat{l}}} = \frac{1}{\Delta \lambda_{\hat{l}}} \int \cdot \, \sin{\zetahat} \, \diff \lambda_{\hat{m}},
\end{align*}
where, $\{a,b,c\} = \{1,2,3\}$ are distinct, and similarly, $\{{\hat l}, {\hat m}\} = \{ {\hat \zeta}, {\hat \psi} \}$ are distinct; $\lambda_{\hat \zeta} = \hat \zeta$ and $\lambda_{\hat \psi} = \hat \psi$.

Under the assumption of gray transport (i.e. $\Ihat = \int_{0}^{\infty}\Ihat_{\hat{\nu}}\diff\hat{\nu}$), this equation simplifies to:
\begin{multline}
    \frac{1}{\Delta t} \left[ \langle n^0 {n}_0 \hat{I} \rangle_{V\hat{\Omega}} \right]_{x_0^-}^{x_0^+}
    + \sum_a \frac{1}{\Delta V} \left[ \Delta A_a \langle {n}^a {n}_0 \hat{I} \rangle_{t A_a \hat{\Omega}} \right]_{x_a^-}^{x_a^+}
    \\ + \sum_{\hat{l}} \frac{1}{\Delta \hat{\Omega}} \left[ \Delta \lambda_{\hat{l}} \langle r^{\hat{l}} {n}_0 \hat{I} \rangle_{t V \lambda_{\hat{l}}} \right]_{\lambda_{\hat{l}}^-}^{\lambda_{\hat{l}}^+}
    = \langle {n}_0 (\hat{J} - \hat{\alpha} \hat{I}) \rangle_{t V \hat{\Omega}}.
    \label{eq:radiative_transfer_discrete}
\end{multline}
In this paper, we focus on the numerical solution of this last equation in a GPU-based framework, and will present
results of frequency-dependent radiative transfer in a future work.

\subsection{Spatial Discretization \label{sec:spatial_discretization}}

\cuharm\ uses a uniform logical Cartesian mesh, whose internal coordinates, $(\tilde{x}, \tilde{y}, \tilde{z})$, are mapped onto any underlying spatial physical coordinates. The mesh boundaries may vary depending on the problem and the underlying physical coordinates. By implementing a non-linear transformation between the logical and physical coordinates, we can achieve a non-uniform, and therefore enhanced, resolution in specific locations of space (e.g., near the BH or close to the equator of an accretion disk). 
%
%

When referring to the resolution in the rest of this document, we will specify the number of cells in each direction as a product corresponding to $N_{\tilde{x}} \times N_{\tilde{y}} \times N_{\tilde{z}}$. All quantities, radiation or MHD, are expressed at the center of each spatial cell and radiation quantities are also understood to be expressed at the center of each angular cell (see Section~\ref{sec:geodesic grid}).

\subsection{Geodesic Grid}
\label{sec:geodesic grid}

The radiation field at each point in space (i.e. in each spatial grid cell) is described on a separate geodesic grid \citep[see e.g.][]{heikes1995numerical, RHR00, FGB13}.
To each spatial cell on the MHD grid, we associate a separate grid describing the direction distribution of the radiation field in that cell.
This geodesic grid is generated by first inscribing an icosahedron on the unit sphere --- the twelve points that make up the icosahedron
are normalized such that they lie on the unit sphere. 

Higher resolutions (otherwise called higher grid generations) can be constructed by successive refinement of this initial 0$^{th}$ generation grid. To carry out this refinement, we first form the Delaunay triangulation by taking each of these twelve points and connecting it to its closest five neighbors, resulting in an icosahedron composed of 20 triangular faces. Next, the sides of each triangular face are bisected, to give a new vertex point at the center of each side. These new vertex points can then be projected onto the unit sphere and connected to their closest six neighbors, to form a new Delaunay triangulation of the unit sphere.

\begin{figure}[htbp!]
    \centering
    \includegraphics[width=0.85\linewidth]{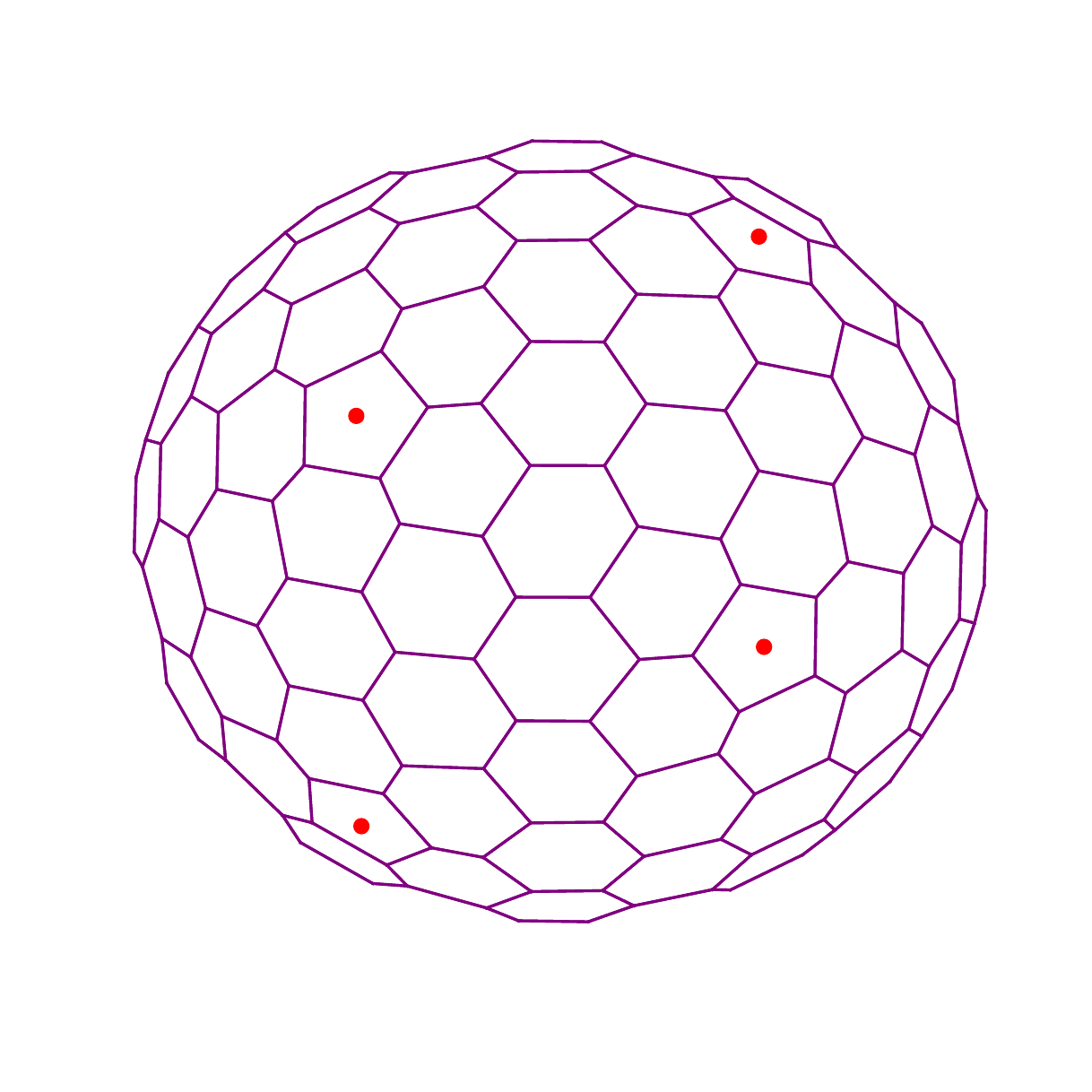}
    \caption{Example of a geodesic grid, here at generation 2, corresponding to 162 directions/angular cells. Only half of the grid is shown, i.e. the directions in the opposite hemisphere are not displayed. The red dots represent vertexes of the underlying icosahedron, which go on to become the centers of the pentagonal cells that make up the 0$^{th}$ order grid. The hexagonal cells are then formed from the refinement process described in Section~\ref{sec:geodesic grid}.}
    \label{fig:geodesic_example}
\end{figure}

To create the final grid used in our calculation, we take the above Delaunay triangulation and form its associated Voronoi diagram.
This Voronoi diagram is formed by taking each central vertex point on the Delaunay triangulation grid to be the center of a grid face.
The boundaries of the Voronoi cells are then defined such that every point inside the cell is closer to that cell's central vertex
than to any other vertex. Each vertex on the original Delaunay triangulation grid becomes the center of a face on the corresponding
Voronoi diagram. The operation is repeated to reach the targeted generation. For generation $s$, this results in a grid
with $N_s = 2 + 10 \times 2^{2 s}$ faces, where the initial twelve faces from the generation 0 grid are pentagonal, while the
remainder of the faces formed from higher grid generations are hexagonal. Throughout the manuscript, we will refer to the geodesic
grid by its generation, where G0 is the non-refined grid, G1 corresponds to the first level of refinement, G2 to the second, etc.
The number of angular cells in the six first grid generations used in this paper are: 12, 42, 162, 642, 2562 and 10242. We show in
Figure~\ref{fig:geodesic_example} a generation 2 geodesic grid.

The geodesic grid presents several advantages over the conventional latitude-longitude grid. Firstly, the grid is made up of an isotropic
set of directions, which are associated to cells of nearly homogeneous solid angle, with variation in area as small as 10\% between the smallest and largest cells. Secondly, to account
for strong anisotropies of the radiation because of the Lorentz boost, the grid can be deformed in advance by performing a Lorentz
transformation of its directions. It can also be easily oriented along a pre-defined direction to enhance numerical accuracy.
In addition, the large number of boundary faces (5 or 6) in each angular cell reduces numerical diffusion and enables precise radiation
transport in curvilinear coordinates.

This grid also presents some disadvantages: it requires more complicated bookkeeping
to account for the multiple neighbors of each angular cell. In addition, our choice of grid refinement technique strongly limits the
flexibility in choosing the number of directions to be used. Finally, the distributions of the photon direction do not present any symmetry which
could be exploited to perform integration of the specific intensity to calculate the moments of the radiation field, e.g. the energy
in the radiation, such as the one presented in \citet{Bruls1999, Jia21} and \citet{MPJ25}. In other words, our discretization does not imply that when the radiation field is isotropic, the radiation flux is
exactly null and that the Eddington tensor is exactly one third of the unit matrix. However, we do not deem this a critical flaw,
since in the relativistic regime considered here, the specific intensity is typically not isotropic in the comoving frame,
while the transport is performed in the tetrad frame.

\subsection{Reconstruction of the Specific Intensity at Spatial and Angular Cell Boundaries \label{sec:reconstruction}}

In order to calculate reconstructed values of $\Ihat_{\hat \nu}$ at the spatial cell boundaries, we use the same method as for the MHD primitives, namely piecewise linear interpolation (PLI). Each angular cell is treated separately, and so we must be careful to use sufficient spatial resolution such that no significant rotation occurs in the tetrad frame. Without a high enough resolution, the rotation of the tetrad frame in space would lead to angular cells pointing in sufficiently different directions that radiation would begin to deviate from the correct trajectory. This effect is significant in strongly curved regions of spacetime, namely close to the event horizon of the central black hole, or when using spherical-like coordinates. In practice, we find that this requirement is satisfied as long as we are using a spatial resolution that is sufficient for a standard, non-radiative MHD problem.

The PLI method cannot be used to reconstruct the value of $\hat I_{\hat \nu}$ at the boundaries of the angular grid. Instead,
the reconstruction between angular cells on the geodesic grid is carried out using the minimum-angle plane reconstruction
method \citep[MAPR;][]{CP08}, which, at least formally, provides second-order accuracy. This reconstruction technique
has also been used by \textit{e.g.} \citet{FGB13} in the context of magnetohydrodynamic simulation (no GR). 
We chose this method over the alternative donor cell method, as used, e.g., in \citet{ZSM25} as this method provides only first order accuracy.



With the reconstructed values at both the spatial and angular boundaries in hand, the fluxes at these boundaries can be calculated.

\subsection{Fluxes \label{sec:fluxes}}

Due to the curvature of space-time and/or choice of coordinates, the tetrad frame changes its orientation on the path of the radiation field. This is described by the third and fourth terms in Equation~\eqref{eq:covariant_radiative_transfer}.  Those two terms therefore describe a spatially localized non-zero flux in the angular grid. 
This change is accounted for by the symbols $r^{\hat \zeta}$, $r^{\hat \psi}$.

While the third and fourth terms in Equation~\eqref{eq:covariant_radiative_transfer} are specialized to the polar and azimuthal coordinates in momentum space ($\hat \zeta$ and $\hat \psi$), our decision to use a geodesic grid prevents us from using this equation directly. Instead, we calculate the flux across each individual edge that separates the hexagons and pentagons which form the geodesic grid. This is achieved by defining a unit vector $\hat u$, that is normal both to the edge and to the geodesic grid, oriented outwards from the cell. 

In addition, for every two angles $\{\hat \zeta,{\hat \psi}\}$ we define two vectors, $u^{\hat \zeta} = \partial_{\hat \zeta}$ and $u^{\hat \psi} = \partial_{\hat \psi}$, calculate the scalar product between these two vectors and $\hat u$, and weight the contribution of the third and fourth terms in Equation~\eqref{eq:covariant_radiative_transfer} by the respective scalar product. 
Therefore, 
we write their total contribution to the angular flux across each boundary of the geodesic grid as  
\begin{equation}
\begin{array} {lcl}
    F_{l,s + \frac{1}{2}} & = & \sum_{\hat j} n_{0, l, s+\frac{1}{2}} r^{\hat j}_{l, s+\frac{1}{2}} \left( \hat u \cdot \hat u^{\hat j} \right) \\& & \times 
    \begin{cases}
        \Ihat_{\mathrm{L},l, s+\frac{1}{2}},& r^{\hat j}_{l, s+\frac{1}{2}} \left( \hat u \cdot \hat u^{\hat j} \right)> 0, \\
        \Ihat_{\mathrm{R}, l, s+\frac{1}{2}},& r^{\hat j}_{l, s+\frac{1}{2}} \left( \hat u \cdot \hat u^{\hat j} \right) < 0,
    \end{cases}
    \label{eq:angular_fluxes}
    \end{array}
\end{equation}
where $\hat j \in \{ \hat \zeta, \hat \psi \}$ and $r^{\hat j}_{l, s+\frac{1}{2}}$ are given by Equation~\eqref{eq:angular_n_vectors}, and calculated at the center of each edge. In addition, $\hat I_\mathrm{L}$ and $\hat I_\mathrm{R}$ refer to the reconstructed values of the intensity on the left and right sides of the grid cell boundaries respectively, as explained in Section~\ref{sec:reconstruction}. This form of the flux ensures that the scheme describes propagation of radiation in the correct direction. 


For the spatial fluxes (between neighboring spatial cells), we discriminate between vacuum and non-vacuum cases. In vacuum, the spatial flux in direction $i \in \{1,2,3\}$ is given by:
\begin{equation}
    F_{i, l+\frac{1}{2}, s} = n_{0, l+ \frac{1}{2}, s} n^{i}_{l+ \frac{1}{2}, s}
    \begin{cases}
        \Ihat_{\mathrm{L}, l+ \frac{1}{2}, s},& n^{i}_{l+ \frac{1}{2}} > 0, \\
        \Ihat_{\mathrm{R}, l+ \frac{1}{2}, s},& n^{i}_{l+ \frac{1}{2}} < 0.
    \end{cases}
    \label{eq:spatial_fluxes_vacuum}
\end{equation}
Note that here $\Ihat_{\mathrm{L}}$ and $\Ihat_{\mathrm{R}}$ correspond to the intensities to the left and right of the spatial boundary, not the geodesic cell boundaries as above. 

Both \citet{Jia21} and \citet{WMJ23} point out that using the spatial fluxes given by Equation~\eqref{eq:spatial_fluxes_vacuum} can result in non-physical fast transport of radiation in regions of high opacity. As opacity increases, radiation diffusion and radiation advection compete as the dominant mechanisms for radiation transport.
In the non-vacuum case therefore, following \citet{Jia21}, we define for each direction $s$ on the geodesic grid and at each spatial location $l$, the approximate radiation signal speeds:
\begin{align}
    S^+_{i, l + \frac{1}{2}, s} = \left \{ \begin{aligned}
     & n^i_{l+\frac{1}{2}, s} \sqrt{\frac{1-\exp(-\tau_{i, l+\frac{1}{2}}^2)}{\tau_{i, l+\frac{1}{2}}^2}} &~~~~~& n^i_{l+\frac{1}{2}, s} > 0\\
    -& n^i_{l+\frac{1}{2}, s} \sqrt{\frac{1-\exp(-\tau_{i, l+\frac{1}{2}}^4)}{\tau_{i, l+\frac{1}{2}}^2}} &~~~~~& n^i_{l+\frac{1}{2}, s} < 0
    \end{aligned} \right.
\end{align}
and 
\begin{align}
    S^-_{i, l + \frac{1}{2}, s} = \left \{ \begin{aligned}
     & n^i_{l+\frac{1}{2}, s} \sqrt{\frac{1-\exp(-\tau_{i, l+\frac{1}{2}}^2)}{\tau_{i, l+\frac{1}{2}}^2}} &~~~~~& n^i_{l+\frac{1}{2}, s} < 0\\
    -& n^i_{l+\frac{1}{2}, s} \sqrt{\frac{1-\exp(-\tau_{i, l+\frac{1}{2}}^4)}{\tau_{i, l+\frac{1}{2}}^2}} &~~~~~& n^i_{l+\frac{1}{2}, s} > 0
    \end{aligned} \right. \label{eq:rad_speeds}
\end{align}
where $\tau_{i, l+\frac{1}{2}} = \bar a \hat\alpha_{l+\frac{1}{2}} dt$ is the average of the effective opacity of the spatial cell $l$ and $l+1$ in direction
$i$, such that the opacity $\hat \alpha_{l+\frac{1}{2}} = (\hat \alpha_l + \hat \alpha_{l+1})/2$ and $\hat \alpha_l, \hat \alpha_{l+1}$ are measured at their respective cell centers. For the numerical coefficient $\bar a$, we note that \citet{Jia21} suggests to use a value between 1 and 10. Throughout this work, we choose $\bar a = 1$. 

For the specific intensity, the numerical flux in direction $i \in \{1,2,3\}$ across spatial cell boundary
$l + 1/2$ and angular direction $s$ is therefore given by:
\begin{align}
    & F_{i, l+\frac{1}{2}, s} \\
     =  & ~~  \frac{S^+_{i, l + \frac{1}{2}, s}}{S^{+}_{i, l + \frac{1}{2}, s} - S^{-}_{i, l + \frac{1}{2}, s}} n^i_{l + \frac{1}{2}, s} \hat I_{R, l + \frac{1}{2}, s} \exp \left (-\tau_{i, l + \frac{1}{2}} \right )  \nonumber \\ 
  - & \frac{S^{-}_{i, l + \frac{1}{2}, s}}{S^{+}_{i, l + \frac{1}{2}, s} - S^{-}_{i, l + \frac{1}{2}, s}} n^i_{l + \frac{1}{2}, s}  \hat I_{L, l + \frac{1}{2}, s} \exp \left ( -\tau_{i, l + \frac{1}{2}, s} \right )  \label{eq:flux_Jiang} \nonumber \\  
  + & \frac{S^{+}_{i, l + \frac{1}{2}, s} S^{-}_{i, l + \frac{1}{2}, s}}{S^{+}_{i, l + \frac{1}{2}, s} - S^{-}_{i, l + \frac{1}{2}, s}} n^i_{l + \frac{1}{2}, s} \left [ \hat I_{ l + 1, s} - \hat I_{l, s} \right ]  \nonumber  \\
  + & \left [ 1- \exp \left (-\tau_{i, l + \frac{1}{2}, s} \right ) \right ] v^i_{R/L, l+\frac{1}{2}} \hat I_{R/L, l+\frac{1}{2},s }. \nonumber 
\end{align}
Here, $v^i_{R/L, l+\frac{1}{2}}$ is the 3-velocity of the flow to the right / left of the boundary between spatial cells $l$ and $l+1$, and the factor $\hat I_{R/L, l+\frac{1}{2},s }$ is chosen to be
$\hat I_{R, l+\frac{1}{2},s }$ if $v^i_{l+\frac{1}{2}} > 0$, and $\hat I_{L, l+\frac{1}{2},s }$ otherwise. If the reconstructed velocities on the right and left of the cell boundary have different signs, we enhance the stability by upwind splitting using one sided speeds, such that $v^i_{R/L, l+\frac{1}{2}} \hat I_{R/L, l+\frac{1}{2}} \equiv v^i_{R, l+\frac{1}{2}} \hat I_{R, l+\frac{1}{2}} + v^i_{L, l+\frac{1}{2}} \hat I_{L, l+\frac{1}{2},s}$. Finally,  $\hat I_{l,s}$, which appears in the third term, is the intensity measured at the center of the cell.


Compared to the fluxes defined by \citet{Jia21}, the first two terms are exponentially cut at large opacity,
which guarantees that they do not contribute in this limit and that the diffusion regime will be reached on the geodesic grid (see Appendix~\ref{app:fluxes} for details). The third term is responsible for radiation diffusion at large opacity. Finally, the last term ensures that
radiation is advected with the flow in regions of high opacity \citep[see e.g.][]{OT16,ATO20}. 

In addition to the above, we impose an important restriction on the fluxes, namely that flux cannot flow between cells
where the sign of $n_0$ changes. This situation can occur in the ergosphere of a rotating black hole, or below
the horizon for a non-rotating black hole and arises because the directions in the tetrad frame can correspond to
both increasing (positive $\hat n^0$) or decreasing (negative $\hat n^0$) time coordinate. The restriction we use is a direct result of applying the geodesic equation:
\begin{equation}
    \frac{\diff k^\alpha}{\diff \lambda} = - \Gamma^{\alpha}_{\beta \gamma} k^\beta k^\gamma
\end{equation}
in a stationary spacetime, and noting that the propagation vector is such that $n_\alpha =  k_\alpha / \nu$, where $k_\alpha$ is the photon four-momentum.
This issue was already reported by \citet{WMJ23}. In their approach, $\Ihat_\nu$ is zeroed
whenever $|n^0|< 0.1$. This prevents exponential growth of the specific intensity as $n^0$ changes sign across spatial cell boundaries. It also avoids radiation energy density build up within the ergosphere. Indeed, we found that keeping angular cells with small $|n^0|$ and only zeroing the specific intensity when $n^0$ changes sign leads to radiation stacking up in this region. This is because radiation flux is directly proportional to $n^0$, and so a small $n^0$ therefore corresponds to very small flux and potential radiation buildup.

\subsection{Radiation-Fluid Interaction \label{sec:interaction}}


The coupling of the radiation sector to the GRMHD sector of the code is achieved by applying  Equation~\eqref{eq:energy_momentum_conservation}. In order to carry out this calculation, we must calculate the radiation four-force, $G_\alpha$, acting over the course of the timestep. We carry out each of the following calculations in a frame that is comoving with the fluid at each point, otherwise known as the fluid rest-frame (see Section~\ref{sec:comoving_frame}). 

To solve the RTE (Equation~\eqref{eq:covariant_radiative_transfer}), we use an operator splitting technique. We independently evolve the transport (left hand side of Equation~\eqref{eq:covariant_radiative_transfer}) and the interaction (right hand side of that equation) terms. Here we specify how the interaction term is calculated. We provide in Section~\ref{sec:temporal} the details of the combined transport and interaction evolution needed to perform a complete time step calculation.



\subsubsection{Mass Scaling\label{sec:mass_scaling}} 

Calculation of the interaction terms requires us to first set the unit scale of our simulation. This scale is naturally induced
by the black hole mass $M$, along with the accretion rate, $\dot{M}$, which sets the density scale. In particular we have:
\begin{multline}
\begin{gathered}
    l_\mathrm{unit} = \frac{GM}{c^2}, \quad t_\mathrm{unit} = \frac{l_\mathrm{unit}}{c}, \quad M_\mathrm{unit} = \dot{M} t_\mathrm{unit}, \\ 
    \rho_\mathrm{unit} = \frac{M_\mathrm{unit}}{t_\mathrm{unit}^3}, \quad u_\mathrm{unit} = \rho_\mathrm{unit}c^2.
\end{gathered}
\label{eq:unit_scale}
\end{multline}
These scales are used to convert between code units and cgs units, which are necessary in the radiation sector,
with $l_{\mathrm{cgs}} = l_{\mathrm{code}}l_{\mathrm{unit}}$, $\rho_{\mathrm{cgs}} = \rho_{\mathrm{code}} \rho_{\mathrm{unit}}$ etc. Here, quantities with subscript ``cgs'' refer to values in cgs units, while subscript ``code'' refers to values in normalized code units.
We note that the mass accretion rate cannot be known in advance, and is provided by experience or normalization
to a previous simulation from which the mass accretion rate can be benchmarked.

\subsubsection{Calculation of the Coupling Terms \label{sec:coupling_calc}}

We begin from the right-hand side of Equation~\eqref{eq:covariant_radiative_transfer}, to write:
\begin{equation}
    \partial_0 \hat{I} = \frac{1}{n^0} \left( \jhat - \alphahat \Ihat\right).
    \label{eq:source_term_tetrad}
\end{equation}
The Lorentz transformations of $\hat I$, $\hat \alpha$ and $\hat j$ allow us to transform this equation to the fluid frame to give:
\begin{equation}
    \partial_0 \bar{I} = \frac{\nbar^{\bar 0}}{n^0} \left( \jbar - \alphabar \Ibar\right),
    \label{eq:source_term_fluid}
\end{equation}
where we have made use of $\bar{\nu}/\hat{\nu} = \bar{n}^{\bar 0}/\nhat^{\hat{0}}$.

This equation represents the change in $\Ibar$ (and consequently $\Ihat$) as a result of the interaction between the radiation field and the plasma with time. Note that throughout the calculation of radiation-fluid coupling, we are implicitly assuming that the fluid velocity does not change over the course of the timestep we are considering. In particular, this assumption is crucial to allow us to transform between Equations \eqref{eq:source_term_tetrad} and \eqref{eq:source_term_fluid}, as we rely on the fact that $\bar{n}^{\bar 0}$ is constant over the timestep for the transform to be valid \citep[see further discussion in ][]{WMJ23}.

In order to proceed, we must define the absorption and scattering opacities, $\kappa_a$ and $\kappa_s$. These opacities are user-defined and problem-dependent, but once specified, allow us to continue with the interaction procedure. Using $\alpha = \rho \kappa$, where $\rho$ is the density and $\alpha$ is the absorption coefficient, we split the emissivity and absorptivity into absorption and scattering terms:
\begin{subequations}
    \begin{align}
        \jbar     &= \jbar^a + \jbar^s, \label{eq:emissivity_total} \\
        \alphabar &= \alphabar^a + \alphabar^s,
    \end{align}
\end{subequations}
where the superscripts $a$ and $s$ refer to absorption and scattering respectively.

We introduce the Planck mean opacity:
\begin{equation}
    \alphabar_\mathrm{P}^a = \frac{4\pi}{\arad \bar{T}^4} \int_0^\infty \alphabar_{\bar \nu} B_{\bar \nu} (\bar{T}) \diff \bar{\nu},
    \label{eq:planck_mean}
\end{equation}
where $\arad$ is the radiation constant, $\bar T$ is the local temperature of the plasma as measured in the comoving frame, and $B_\nu (\bar{T})$ is the Planck function. Using Kirchoff's law, we can then define the emissivity as:
\begin{equation}
    \jbar^a = \frac{1}{4\pi} \alphabar_\mathrm{P}^a \arad \bar{T}^4.
    \label{eq:emissivity_kirchoff}
\end{equation}

For the scattering process, we assume it to be both elastic and isotropic in the fluid-frame, to write
\begin{equation}
    \oint \left( \jbar^s - \alphabar^s \Ibar\right) \diff \bar \Omega = 0.
\end{equation}
This leads directly to an expression for the emissivity due to scattering, given by:
\begin{equation}
    \jbar^s = \frac{1}{4\pi} \alphabar^s \Ebar,
    \label{eq:emissivity_scattering}
\end{equation}
where 
\begin{equation}
\Ebar = \oint \Ibar \diff \bar\Omega
\label{eq:Ebar_calc}
\end{equation}
is the first moment of the specific intensity, representing the (comoving) radiation energy density. 
Using these relations, Equation~\eqref{eq:source_term_fluid} becomes:
\begin{equation}
    \partial_0 \bar{I} = \frac{\nbar^{\bar0}}{n^0} \left( \frac{1}{4\pi} \left( \alphabar_\mathrm{P}^a \arad \bar{T}^4 + \alphabar^s \Ebar\right) - \alphabar \Ibar\right).
    \label{eq:fluid_interaction_exact}
\end{equation}

Following the discussion in \citet{WMJ23}, we choose to adopt an implicit scheme for the solution of Equation~\eqref{eq:fluid_interaction_exact}. This is because radiation interaction takes place on time scales that are typically several orders of magnitude shorter than the dynamical time scale. Applying implicit finite differencing to Equation~\eqref{eq:fluid_interaction_exact}, with subscripts $+$ and $-$ denoting values before and after the coupling step respectively, leads to the following expression for $\Ibar_+$ (the fluid frame intensity after the radiation-fluid interaction):
\begin{align}
    &\Ibar_+ = \frac{n^0}{n^0 + \bar{n}^{\bar 0} \alphabar \Delta t} \Ibar_- \nonumber \\ 
    + &\frac{\bar{n}^{\bar 0} \Delta t}{4\pi \left( n^0 + \bar{n}^{\bar 0} \alphabar \Delta t \right)} \left( \alphabar_{\mathrm{P} }^a \arad \bar{T}_+^4 + \alphabar^s \Ebar_+ \right).
    \label{eq:fluid_interaction_discrete}
\end{align}

In \cuharm\ we implement three different methods to solve Equation~\eqref{eq:fluid_interaction_discrete}, which differ by the implicit or explicit nature of the emission and absorption coefficients.
The first is the method used in \citet{WMJ23}, in which all coefficients are treated explicitly, i.e., depending only on $\bar T_-$. After implementing this method, we found that it is numerically unstable, and leads to negative or undefined temperatures. This is because the emission and absorption coefficients strongly depend on the temperature, such that small variations in the temperature can lead to huge variations in these quantities.
We therefore implemented a second method, which relies on expressing the Planck absorption coefficient $\bar\alpha_{\mathrm{P} }^a$ implicitly, namely $\bar\alpha_{\mathrm{P} }^a (\bar T_+)$, while keeping the total absorption coeffecient $\alphabar (\bar T_-)$ explicit. This does not add to the calculation time, but neglects the strong dependence of $\alphabar$ on the temperature. Finally, we use a third method, where $\alphabar$ is treated semi-implicitly. We describe here all three methods we use.

{\it Method 1.} The first method follows the treatment by \citet{WMJ23}. 
In order to solve Equation~\eqref{eq:fluid_interaction_discrete} for the post-interaction intensity, we require an expression for $\Ebar_+$. Since we do not know $\Ibar_+$ at this point, we cannot use Equation~\eqref{eq:Ebar_calc} for this purpose. We therefore multiply Equation~\eqref{eq:source_term_tetrad} by $n^0 n_\alpha$ and integrate over solid angle. Using Equations \eqref{eq:galpha} and \eqref{eq:R}, this leads to:
\begin{equation}
    \partial_{0}R^{0}_{~\alpha} = -G_\alpha.
\end{equation}

We can then combine this with Equation~\eqref{eq:energy_momentum_conservation},
\begin{equation*}
    \partial_0 \left( T^{0}_{~\alpha}\right) = G_\alpha,
    \label{eq:operator_split_Tmunu}
\end{equation*}
to write:
\begin{equation}
    \partial_0 \left( T^{0}_{~\alpha} + R^{0}_{~\alpha} \right) = 0.
    \label{eq:energy_exchange}
\end{equation}
Since we have assumed that the fluid velocity is unchanged during the interaction phase, we can equivalently write:
\begin{equation}
    \partial_0 \left( \bar{T}^{\bar 0 \bar \alpha} + \bar{R}^{\bar 0 \bar \alpha} \right) = 0.
    \label{eq:energy_exchange_fluid}
\end{equation}

Then, since radiation does not change the plasma density, nor does it interact with magnetic fields directly, and assuming an ideal gas, we can use the fact that $\bar{R}^{\bar 0 \bar 0} = \Ebar$ and $\bar{T}^{\bar 0 \bar 0} = \bar{\rho} + \bar{u}_\mathrm{gas} + \bar{u}_\mathrm{mag}$ to write:
\begin{equation}
    \frac{k_\mathrm{B} \bar{\rho}}{\left( \Gamma - 1 \right) \mu m_\mathrm{p}} \left( \bar{T}_+ - \bar{T}_- \right) = - \left( \Ebar_+ - \Ebar_- \right),
    \label{eq:ideal_gas_difference}
\end{equation}
where $\bar{u}_\mathrm{gas}$ and $\bar{u}_\mathrm{mag}$ are the energy densities of the gas and magnetic fields respectively, as measured in the fluid frame, $k_\mathrm{B}$ is Boltzmann's constant, $\bar{\rho}$ is the mass density, $\Gamma$ is the adiabatic index of the plasma, $\mu$ is the mean molecular weight of the plasma, and $m_\mathrm{p}$ is the mass of the proton. In essence, Equation~\eqref{eq:ideal_gas_difference} states that the energy gained (lost) by the radiation is exactly the energy lost (gained) by the plasma.

We next integrate Equation~\eqref{eq:fluid_interaction_discrete} over solid angle, giving:
\begin{equation}
    \Ebar_+ = A_1 + A_2 \left( \alphabar_\mathrm{P}^a \arad \bar{T}_+^4 + \alphabar^s \Ebar_+\right),
    \label{eq:energy_integral}
\end{equation}
where we define
\begin{subequations}
    \begin{align}
        A_1 &= \oint \frac{n^0}{n^0 + \bar{n}^{\bar 0} \alphabar_- \Delta t} \Ibar_- \Delta \bar{\Omega}, \label{eq:A1}\\
        A_2 &= \frac{\Delta t}{4 \pi} \oint \frac{\bar{n}^{\bar 0}}{n^0 + \bar{n}^{\bar 0} \alphabar_- \Delta t} \Delta \bar{\Omega},\label{eq:A2}
    \end{align}
\end{subequations}
where we emphasize again that $\bar \alpha$ is calculated explicitly. This removes the requirement to recalculate
$A_1$ and $A_2$ in the iterative solver for the temperature, avoiding the costly angular integration performed in Equations
\eqref{eq:A1} and \eqref{eq:A2} when solving Equation~\eqref{eq:implicit_T} below numerically.

Under the assumption that there is no radiative process that changes the gas density, and knowing the value of $\Ebar_-$ via Equation~\eqref{eq:Ebar_calc}, all that is now required is an expression for $\bar{T}_+$, which allows us to calculate the change in temperature due to radiation, and therefore the change of energy. Combining Equations \eqref{eq:ideal_gas_difference} and \eqref{eq:energy_integral} results in the following quartic equation for $\bar{T}_+$:
\begin{align}
\begin{aligned}
    & A_2 \bar \alpha_p \arad \bar{T}^4_+ + Y (1 - A_2 \bar \alpha^s) \bar{T}_+ \\
    + &A_1 - Y (1 - A_2 \bar \alpha^s) \bar{T}_-  - (1- A_2 \bar \alpha^s) E_- = 0,
\end{aligned}  \label{eq:semi_implicit_T}
\end{align}
where we define $Y \equiv (k_B \bar{\rho})/[\mu m_p (\Gamma -1 )]$. We solve this equation numerically using
the Newton-Raphson method, as the derivative of this expression is readily obtained.

{\it Method 2.} We find that this semi-implicit method, in the sense that the emission coefficient and absorption are estimated
using $\bar{T}_-$, can be numerically unstable and does not guarantee the existence of a positive root for the temperature. In
astrophysical calculations, this instability originates from the strong dependence of the emission coefficients on the
temperature. For the bremsstrahlung opacity used in the Section~\ref{sec:results} below for example, the dependence is $\bar \alpha^a_E \propto \bar{T}^{-7/2}$
and $\bar \alpha^a_P \propto \bar{T}^{-7/2}$. Therefore even a small variation of temperature can lead to a dramatic variation
of these coefficients. We therefore propose two new alternatives, as explained above.

The second method we use is similar to the original \citet{WMJ23} method, with the exception of expressing the absorption coefficient $\bar \alpha_P^a$ implicitly, such that
$\bar \alpha_P^a = 8 \times 10^{22} \bar{\rho}^2 \bar{T}_+^{-7/2} \equiv \bar N \bar{T}_+^{-7/2} $ (the numerical coefficient is relevant for pure hydrogen). We do not modify
$A_1$ and $A_2$, which are still calculated using explicit emission coefficients, that is, depending on $\bar{T}_-$ only.
This leads to the following equation, which we again solve numerically, using the Newton-Raphson method:
\begin{align}
\begin{aligned}
    & A_2 \bar N \arad \sqrt{\bar{T}_+} + Y (1 - A_2 \bar \alpha^s) \bar{T}_+ + A_1  \\
    - & Y (1 - A_2 \bar \alpha^s) \bar{T}_- - (1- A_2 \bar \alpha^s) E_-  = 0
\end{aligned}  \label{eq:implicit_T}
\end{align}

{\it Method 3.} We find that there exists a region of parameter space, characterized by a high density and a low temperature, in which
calculating $A_1$ and $A_2$ with the emission coefficients expressed as a function of $\bar{T}_-$ leads to inconsistency
in the numerical method. To avoid this issue, we implement our third (semi-implicit)  method: we rewrite the coefficients $A_1$ and $A_2$ as:
\begin{subequations}
    \begin{align}
        A_1 &= \frac{1}{\alphabar \Delta t} \oint \frac{n^0}{ \frac{n^0}{\alphabar \Delta t} + \bar{n}^{\bar 0} } \Ibar_- \Delta \bar{\Omega}, \\
        A_2 &= \frac{1}{\alphabar \Delta t} \frac{\Delta t}{4 \pi} \oint \frac{\bar{n}^{\bar 0}}{ \frac{n^0}{\alphabar \Delta t} + \bar{n}^{\bar 0} } \Delta \bar{\Omega}.
    \end{align}
\end{subequations}
We then use an implicit version of the emission coefficients for the term outside of the integral
and an explicit one for the emission coefficients inside the integral, such that new expressions for $A_1$ and $A_2$ become:
\begin{subequations}
    \begin{align}
        A_1^\prime  &= \frac{1}{\alphabar_+ \Delta t} \oint \frac{n^0}{ \frac{n^0}{\alphabar_- \Delta t} + \bar{n}^{\bar 0} } \Ibar_- \Delta \bar{\Omega}, \\
        A_2^\prime &= \frac{1}{\alphabar_+ \Delta t} \frac{\Delta t}{4 \pi} \oint \frac{\bar{n}^{\bar 0}}{ \frac{n^0}{\alphabar_- \Delta t} + \bar{n}^{\bar 0} } \Delta \bar{\Omega},
    \end{align}
\end{subequations}
where the subscripts $+$ and $-$ added to $\alphabar$ respectively represent a quantity calculated with $\bar{T}_+$ or $\bar{T}_-$.
By doing this, the angular integrals need only be calculated once. These new expressions lead to the following equation, to be solved for $\bar{T}_+$:

\begin{align}
\begin{aligned}
    &Y \alphabar^s \left(A_2^\prime  -  \Delta t \right) \bar{T}_+^{4.5} - A_2^\prime \bar N \arad \bar{T}_+^4 \\
    - &\left [  A_2^\prime \alphabar^s \left( \Ebar_-  + Y \bar{T}_- \right) + A_1^\prime - \left( \Ebar_-  + Y \bar{T}_- \right) \alphabar^s  \Delta t \right ] \bar{T}_+^{3.5} \\
    - & Y \underline \alpha^a \Delta t \bar{T}_+ + \left( \Ebar_-  + Y \bar{T}_- \right) \underline \alpha^a \Delta t = 0.
    \label{eq:T_almost_fully_implicit}
\end{aligned}
\end{align}
where we used the temperature-independent quantity, $\underline \alpha^a$ defined via $\alphabar^a \equiv \underline \alpha^a T^{-7/2}$.
We solve this equation using the Newton-Raphson method, since the derivative is readily obtained.

Once $\bar{T}_+$ is known, we are in a position to calculate $\Ebar_+$ from Equation~\eqref{eq:ideal_gas_difference}. Using this, we
update the value of the specific intensity, $\Ibar_+$, as a result of the radiation-fluid interaction. This new value of $\Ibar_+$
is then transformed back into the tetrad frame for use in the transport sector. If either method 2 or method 3 is used to calculate the temperature, the opacities must be updated and are therefore recalculated for the new value of $\bar T^+$.


\subsubsection{Calculation of The Radiation 4-Force}

With the temperature implicitly estimated, the radiation can be updated via Equation~\eqref{eq:fluid_interaction_discrete}. The radiation 4-force can then be estimated in two different ways. Formally, 
it can be calculated in the tetrad frame as
\begin{align}
    G^{\alphahat} = -(\alphabar^a_P a_{\rm rad} T^4_+ + \alphabar^s_E \bar E_+) u^{\alphahat } - \alphabar_E u_{\hat \beta} R^{\alphahat \hat \beta},
\end{align}
where the radiation stress energy tensor is calculated with the recently updated intensity. However, using
this formula makes the code non conservative due to numerical discretization. We instead follow \citet{WMJ23} and calculate the radiation 4-force
as the temporal variation of $R^{\hat \alpha \hat \beta}$, namely
\begin{align}
    G^{\hat \alpha } = \frac{R^{\hat 0 \hat \alpha}_{t + \Delta t} - R^{\hat 0 \hat \alpha}_{t}}{\Delta t}
\end{align}
After transforming the radiation 4-force to the coordinate frame, it is applied as-is to the right-hand side of
Equation~\eqref{eq:energy_momentum_conservation}.

\subsection{Temporal Update}
\label{sec:temporal}

The hydrodynamical sector is evolved using predictor-corrector time-stepping, as explained in \citet{BPZ23}. We keep this scheme in place
for the hydrodynamical sector and graft the radiation onto it. Each time step is calculated as follows:
start by calculating all fluxes for both the hydrodynamical
and radiation sectors. With the calculated fluxes, update the radiation explicitly by half a time step to account for the radiation
transport. Then, the local interaction is calculated implicitly, resulting in an updated specific energy and the radiation
4-force $G_\alpha$. The radiation 4-force is then used as a source term in the magnetohydrodynamical update. The corrector
step proceeds similarly, but instead using quantities calculated from the first half time step to calculate fluxes and interactions.

The numerical scheme we use can thus be summarized as follows:  
\begin{enumerate}
    \itemsep0em 
    \item Calculate fluxes (both MHD and radiation).
    \item Apply radiation fluxes explicitly to calculate the specific intensity after transport. 
    \item Perform radiation-fluid interaction implicitly.
    \item Using the radiation four-force calculated in the previous step, perform the MHD update.
    \item Repeat for the corrector step.
\end{enumerate}
We find this scheme to provide a good balance between stability and computational cost. 


\subsection{Implementation Notes}

Here, we provide some details about our implementation and describe some challenges we addressed when implementing
the radiation sector.

\textit{Time step:} The MHD sector implements a time-varying time stepping procedure, in which the timestep is set as a
fraction of the time it takes for the fastest MHD wave to cross any of the cells in any direction. When
radiation is enabled, the time step is additionally bounded to half of the light crossing time of the smallest cell.
In this way, we ensure that we capture the radiation free streaming regime. In principle, we could have been less
restrictive on the timestep since in region of moderate to high opacity the ``radiation speed'' is reduced. However,
in the context of accretion, there is always an empty region above the disk whose opacity is small, which would
maintain conditions close to the free streaming limit for the time step.

\textit{Floor model:} Most MHD schemes implement floors to maintain stability of the numerical evolution, see
\textit{e.g.} \citet{MG04,MTB12, PCN19}. In \cuharm\, we use floors in magnetically dominated regions, as 
well as in gas pressure dominated regions by maintaining a minimum values of the ratios $\rho/b^2$, $u_g/b^2$, and $u/\rho$.
However, when radiation is enabled, maintaining these floors leads to catastrophic failures. For instance, in region in
which radiation dominates the energy budget, the internal energy of the gas can be very low, while the density is substantial.
Bounding the ratio $u / \rho$ leads to over-estimating the gas temperature, which provides large amounts of energy to the radiation
field, ultimately destroying the solution. 

To maintain numerical stability, we use the following floors and limits: (i) When
the density (in code units) is smaller than $5 \times 10^{-4}$, radiation interaction is not considered, the emission and
absorption coefficient are set to 0, and radiation is allowed to free stream in these regions. (ii) The ratios $u / \rho$ and $u / b^2$
are no longer bounded, and $u$ can become small in regions dominated by radiation. The energy equation can become unsuitable to
solve for $u$. After the conserved to primitive variable inversion, we estimate the temperature. If the temperature is
different by a factor more than 3 compared to the temperature estimated in the interaction step, then
we reset $u_g$ to the value it would have if the temperature were the same after the interaction step.   

\section{Tests \label{sec:tests}}

\subsection{Spherical Light Expansion \label{sec:spherical_expansion}} 

As an initial test of the basic functionality of our transport scheme, we carry out a ``spherical expansion'' test in 3 dimensions,
in which a small point source, measuring 3 spatial cells in each direction, is initialized with an isotropic intensity. The system is then allowed to evolve.
Aside from this initial setup, there is no other source of radiation
and we expect light to propagate outwards, forming a sphere (which appears as a ring in a 2D slice) around the central point. This test
is designed to mimic a single flash of radiation at the center of the simulation domain, which then propagates outwards. The
radiation is initialized at the center of a domain of spatial extent $[-2,2]^3$, discretized
with 64 cells in each direction. For the angular grid, we used a generation 3 geodesic grid (G3)
consisting of 642 angular grid cells.

\begin{figure}[htbp!]
    \plotone{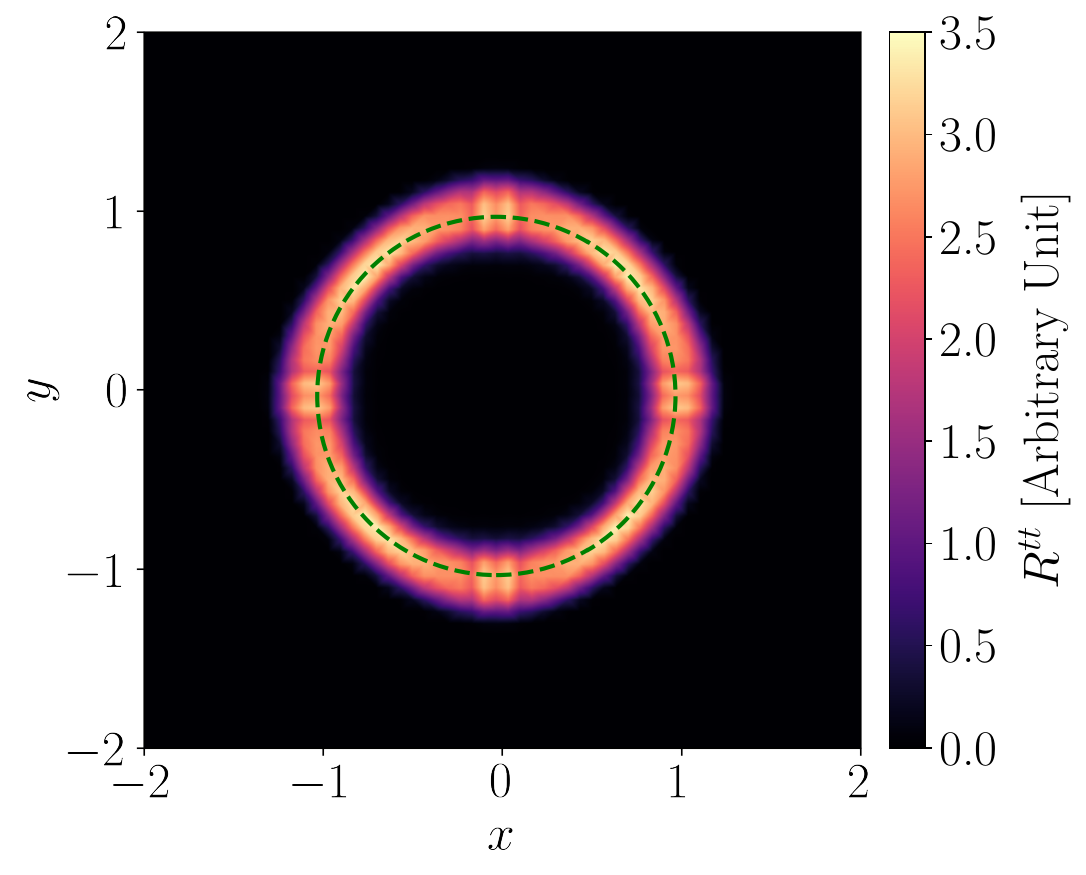}
    \caption{Spherical light expansion test, shown here at $t=1$, demonstrating that our radiation scheme displays the correct behavior, in particular the correct rate of expansion. The green dashed line represents the expected position of photons emitted at the center of the system at $t = 0$. The apparent thickness of the ring at $x = 1$ is due to both numerical diffusion and to the initial setup, which does not exactly model a point source, but rather an extended region of 3 cells in each direction. The limit of the angular resolution becomes visible with variation of the energy density around the ring, in particular along the $x$ and $y$ axes. \label{fig:spherical_expansion}. }
\end{figure}

We show the result of this test in Figure~\ref{fig:spherical_expansion} at time $ t = 1.0$, after which the radiation front is
expected to have moved by a distance $d = 1$ away from the center (recall that we use normalized units, and in particular $c = 1$).
We indeed observe the correct rate of spherical expansion of the radiation throughout the simulation domain, as shown by the
green dashed line at $r=1$.

Some angular discretization effects become visible at later times, becoming more noticeable at large distances or
smaller angular resolution (not demonstrated here). These effects manifest as a separation of the radiation into discrete packets, essentially
as an imprint of the angular grid on the evolution. This is an unavoidable consequence of having a discrete sample of
the angular distribution of the radiation field, rather than having the radiation filling the space continuously. This is clearly visible in Figure~\ref{fig:spherical_expansion} where the radiation energy density is seen to have
an irregular profile.

\subsection{Inverse-Square Decay \label{sec:inverse_square}}

\begin{figure*}[htbp!]
    \plottwo{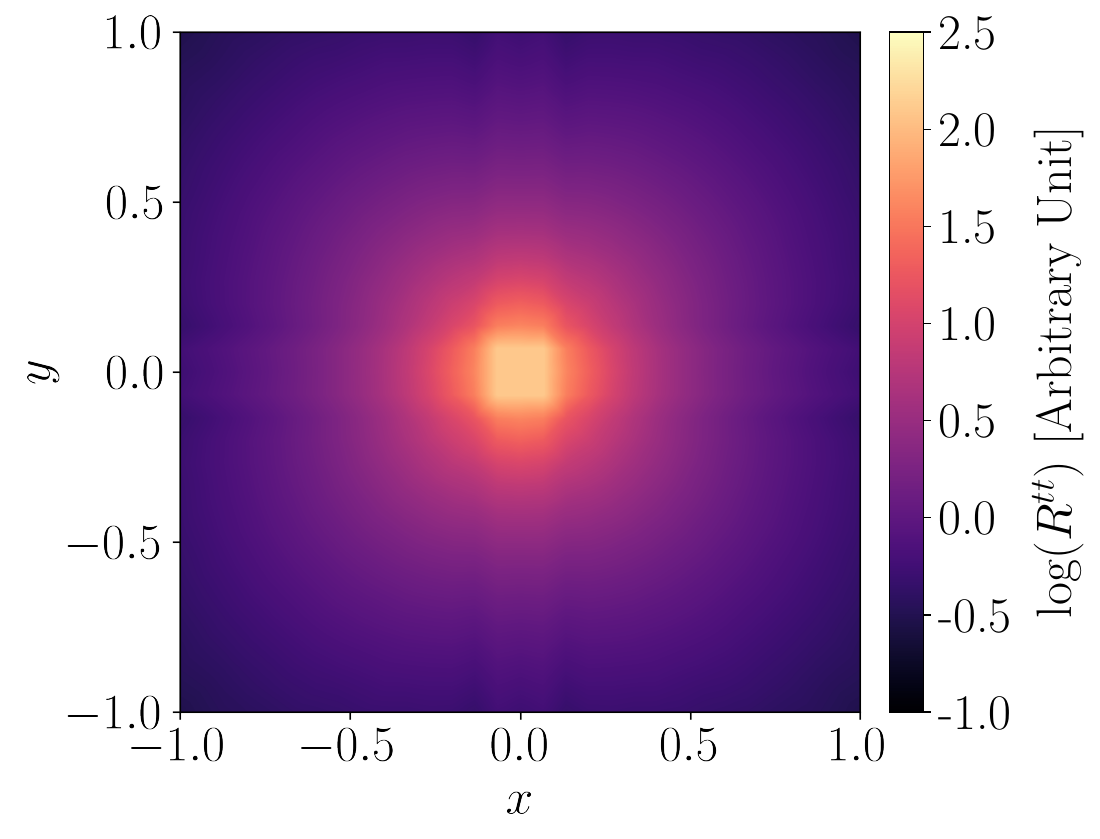}{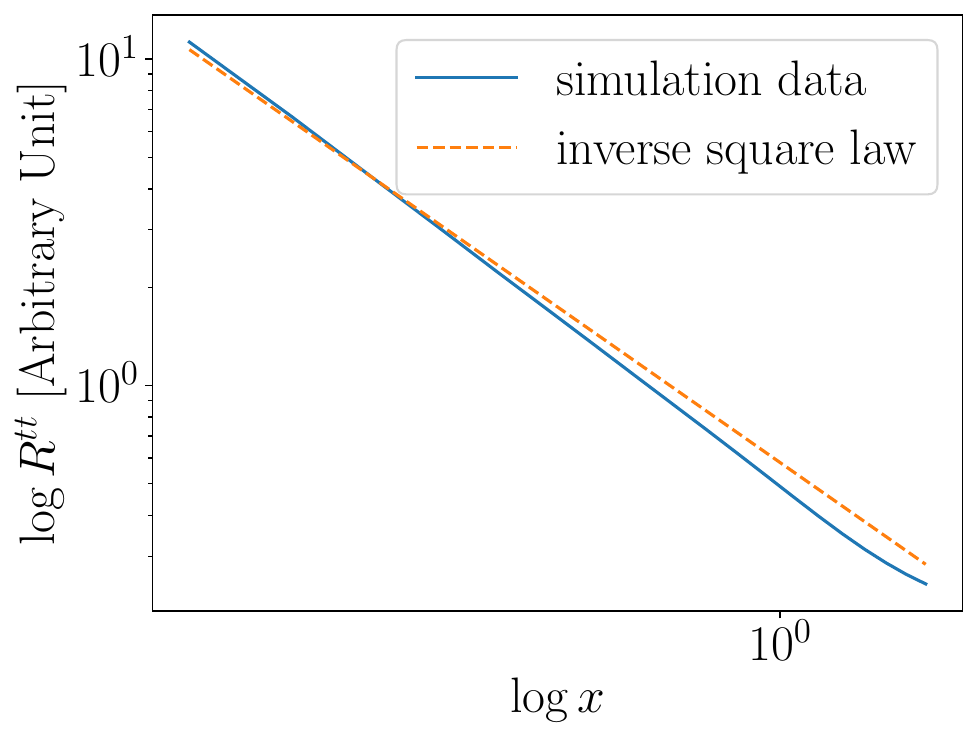}
    \caption{Left. Results of the central source test, which demonstrates that radiation fills in the space nearly uniformly, with the proper decay with distance from the source. Some visible angular effects despite the use of a G3 geodesic grid.
    Right. Plot of radiation energy density $R^{tt}$ along the $+x$ direction in the spherical source test, demonstrating the correct inverse-square decay in intensity with distance from the central source.
    \label{fig:spherical_source}}
\end{figure*}

Following on from the above test, in order to test the quantitative behavior of the outward expansion, we modify the test such
that the point source at the center is now a continuous, steady source, emitting isotropically (as in
Figure~\ref{fig:spherical_source}, left), rather than an instantaneous flash. We allow the system to evolve to long times, where the
radiation front has propagated beyond the edges of our simulation domain. The rate of decay of the radiation energy density with
distance can then be measured, showing a very good agreement with the expected $1/d^2$ decay with distance,
$d$, from the central source (see Figure~\ref{fig:spherical_source}, right).


\subsection{Colliding Beams in Flat Spacetime \label{sec:colliding_beams}}

An important aim of our radiation transport scheme is to accurately capture the angular profile of the
radiation field across the full simulation domain, and as such it is important that our code can correctly handle the propagation
of radiation simultaneously in several directions. We present here a
test of crossing beams. Scenarios such as this are a key differentiator of our chosen method over moment methods such as M1,
in which the beams do not correctly ``pass through'' one another, but rather converge into a single ``average'' radiation field
\citep[see Figure~3 in][for a comparison of the two methods]{ATO20}.

To verify our code's ability to properly capture this behavior, we initialize two beams in a 2D spatial domain of
size $[-4,4]^2$. The radiation field in each beam is initialized as an $8 \times 8$ square in a single direction, i.e., the specific intensity is non-zero in only one angular cell. The directions of the beams are chosen such that they cross
close to the center of the domain. This test was run with a spatial resolution of $128 \times 128$ and a G2 angular grid.

\begin{figure}[htbp!]
    \plotone{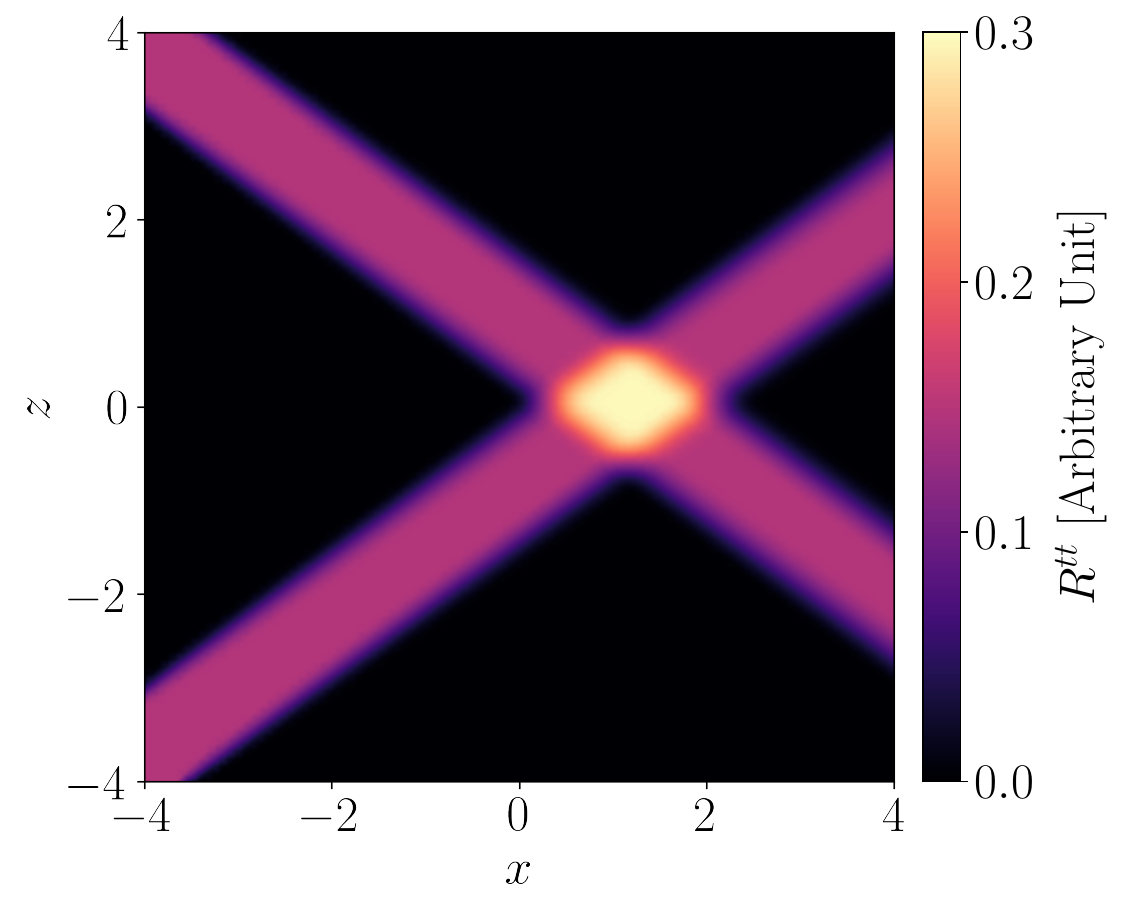}
    \caption{Results of the colliding beam test with resolution $128 \times 128$ and G2 angular grid. The angular structure of the radiation field is maintained, even in the region where there is contribution from both sources. 
    }
    \label{fig:colliding_beams}
\end{figure}

The results of this test are presented in Figure~\ref{fig:colliding_beams}, showing in particular that the beams continue on
their original trajectory after crossing without affecting one another, as well as the correct increase in total
radiation energy density at the intersection of the two beams. This is a clear demonstration that our
implementation correctly accounts for the angular distribution of the radiation field and maintains this distribution accurately during propagation, even in the presence of multiple sources of radiation. 

\subsection{2-Dimensional Hohlraum \label{sec:hohlraum}}

\begin{figure}[htbp!]
    \plotone{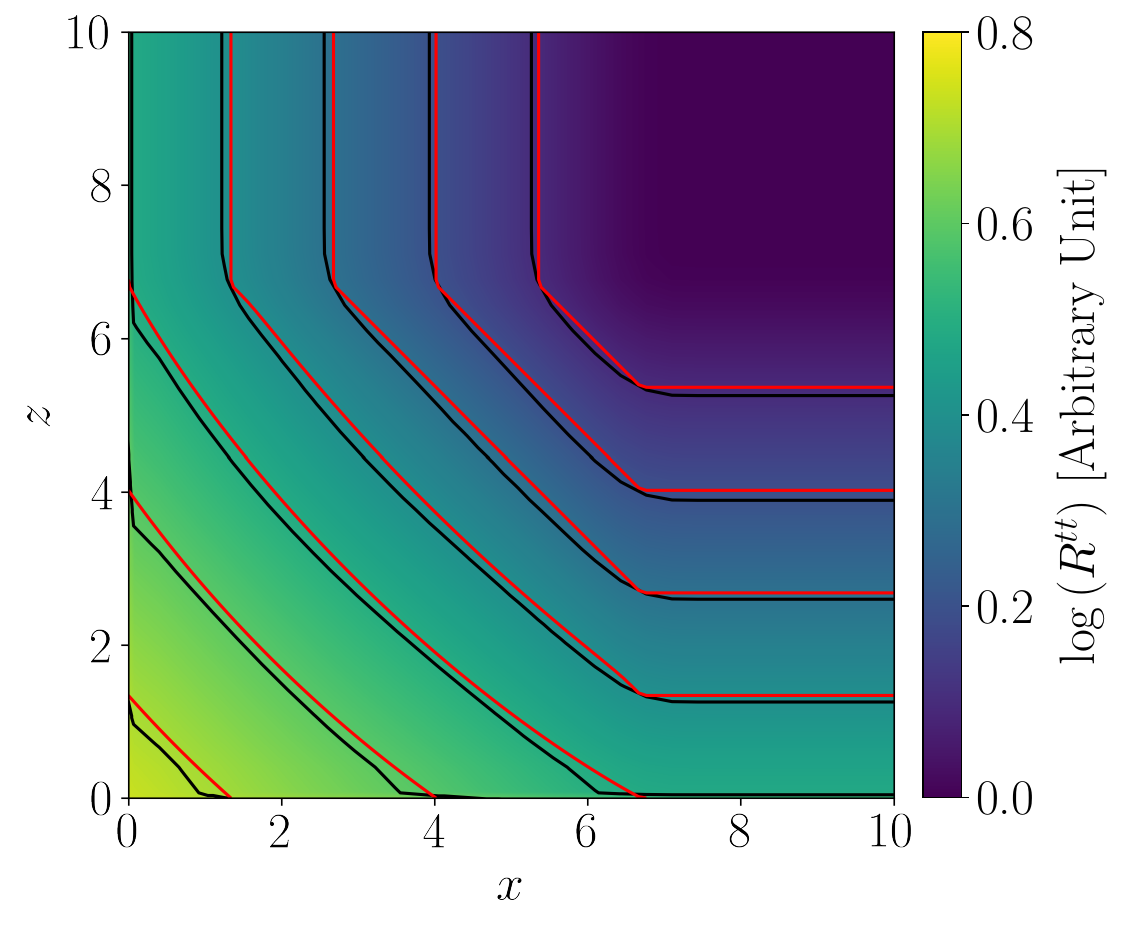}
    \caption{ Results of the 2D hohlraum test, showing the results obtained numerically by our code, with spatial resolution $64 \times 64$ and G4 angular grid. Here only half of the simulation domain in each spatial direction is shown, since the analytic solution assumes that the radiative walls extend to infinity, i.e., neglects the contribution from the domain boundaries. The black lines label contours every 0.1 in arbitrary energy density units for our numerical results, while the red lines represent the same, but for the exact solution, as given by Equation~\eqref{eq:hohlraum_exact}. These results show that our code performs very well at recovering the analytical solution in this test.\label{fig:hohlraum}}
\end{figure}

\citet{RD20} proposed a 2-dimensional ``hohlraum'' test, in which the simulation domain is a vacuum for $x > 0$ and $y > 0$,
and where the boundary walls at $x = 0$ and $y = 0$ are fixed to  radiate isotropically at a constant rate into the vacuum
region, which is sufficiently extended such that edge effects do not affect our analysis. Here,  we extend the simulation domain
$[0, 20]^2$ far beyond the region of interest $[0,10]^2$ shown in Figure~\ref{fig:hohlraum}. We run this test using a spatial
resolution of $64 \times 64$, with a G4 angular grid.

The results are shown in Figure~\ref{fig:hohlraum} at time $t = 6.7$, where we see that our numerical results agree
very closely with the expected analytical result:
\begin{equation}
\begin{cases}
    R^{tt}_x = \frac{1}{2} - \frac{\left( \pi - \theta_x \right) x}{2\pi t} - \frac{1}{2\pi} \sin^{-1} \left( \frac{x \sin{\left( \theta_x \right)}}{\sqrt{x^2 + y^2}}\right), \\
    R^{tt}_y = \frac{1}{2} - \frac{\left( \pi - \theta_y \right) y}{2\pi t} - \frac{1}{2\pi} \sin^{-1} \left( \frac{y \sin{\left( \theta_y \right)}}{\sqrt{y^2 + x^2}}\right),
\end{cases}
    \label{eq:hohlraum_exact}
\end{equation}
where $t$ is the time and $\theta_{x}$ is the first-quadrant angle, defined by:
\begin{equation}
    \cos{\left( \theta_{x} \right)} = \min{\left( \frac{x^2}{\sqrt{t^2 - x^2}}\right)},
\end{equation}
and similarly for $y$. $R^{tt}_x$ and $R^{tt}_y$ are the radiation energy densities at distance $x$ and $y$ from the boundaries.
The total energy density at a point $(x, y)$ is therefore  $R^{tt}_\mathrm{tot} =  R^{tt}_x +  R^{tt}_y$ \citep[see e.g.][for
details of the analytical solution]{RD20, WMJ23}.

This test acts as a further demonstration of the ability of \cuharm\ to resolve the angular distribution of the radiation field
and to correctly evolve it over time, in an environment in which extended radiation sources exist, and radiation is propagating in multiple directions, leading to an anisotropic radiation field. 


\subsection{Comparison of Tetrad Choices \label{sec:tetrad_choice_test}}

In order to calculate the radiative transport, a valid tetrad must be chosen (see Equation~\ref{eq:tetrad_def}). Although formally two different tetrads provide the same results, numerically they lead to different amounts of numerical diffusion, thereby impacting the accuracy of the numerical solution of the radiative transfer equation. Therefore, a proper choice of the tetrad is important to minimize numerical effects. 

\begin{figure*}[htbp!]
    \plotone{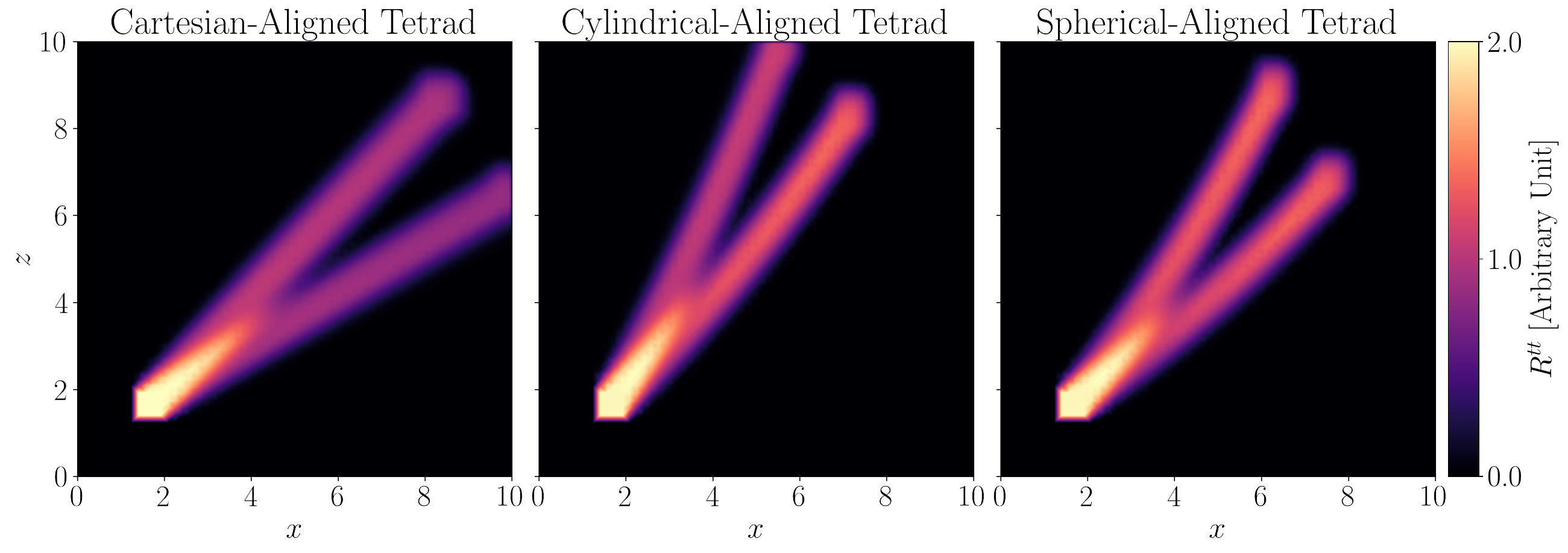}
    \caption{Comparison of three possible tetrad choices in flat spacetime, described by a standard Cartesian coordinate system $(x, y, z)$. In the left-hand panel, the tetrad is aligned directly with the Cartesian unit vectors $(\hat x, \hat y, \hat z)$, and so we recover perfectly straight beam propagation. In the center panel, the tetrad is aligned with the cylindrical unit vectors $(\hat r, \hat \phi, \hat z)$ as described in Equation~\eqref{eq:cylindrical_alignment}, again demonstrating the correct beam behavior, albeit with a slight misalignment. Finally, the right-hand panel shows the results of the same test with a tetrad aligned along the spherical unit vectors $(\hat r, \hat \theta, \hat \phi) $, as in Equation~\eqref{eq:spherical_alignment}. In this case, we note that while the direction of the beams are correct, there is some bending of the rays from the expected direction. \label{fig:tetrad_comparison}}
\end{figure*}

Even in a flat spacetime, depending on the choice of tetrad and coordinates, the Ricci rotation coefficients may be
non-zero, as the tetrad basis itself may rotate as radiation moves across the domain of the simulation.
This situation occurs, for example, in Minkowski space when we use Cartesian coordinates to describe spacetime,
with a tetrad aligned along the spherical unit vectors, $\hat{r}$, $\hat{\theta}$ and $\hat{\phi}$. Thus, the goal of this test
is to show that our implementation can correctly handle multiple tetrad choices for a given spacetime, while also illustrating the varying degrees of numerical diffusion across these different choices.

For this test, we use Minkowski space, described by Cartesian coordinates. We initialize  a source of radiation of size $2 \times 1 \times 2$ spatial cells close to spatial position $\{x,y,z\} = \{2,0,2\}$. For the radiation direction, all angles in the $x-z$ plane within $\pi/6$ of the $\pi/4$ direction are initialized with normalized intensity 2. This corresponds to two angular cells of the geodesic grid having non-zero intensity.

We calculate the propagation of the radiation field from this source. We perform three simulations with this setup, using a different tetrad for each. First, we use a tetrad aligned along the Cartesian unit vectors (i.e. $e^{\hat{t}}_{\delta} = \mathbb{I}$). The results of this case are shown in the leftmost panel of Figure~\ref{fig:tetrad_comparison}. 

Next, we repeat the test for a tetrad aligned along the cylindrical unit vectors $\hat{r}$, $\hat{\phi}$ and $\hat{z}$:
\begin{equation}
    e^{\delta}_{\hat{\alpha}} =
    \begin{pmatrix}
        1 & 0 & 0 & 0 \\
        0 & 0 & 0 & 1\\
        0 & -\cos{\phi} & \sin{\phi} & 0 \\
        0 & \sin{\phi} & \cos{\phi} & 0 
    \end{pmatrix},
    \label{eq:cylindrical_alignment}
\end{equation}
where $\phi = \mathrm{atan2}(y,x)$,  with atan2 being the 2-argument arctangent function.

Finally, we use a tetrad aligned along the spherical unit vectors $\hat{r}$, $\hat{\theta}$ and $\hat{\phi}$:
\begin{equation}
    e^{\delta}_{\hat{\alpha}} =
    \begin{pmatrix}
        1 & 0 & 0 & 0 \\
        0 & \cos{\theta}\sin{\phi} & \cos{\theta}\cos{\theta} & -\sin{\theta} \\
        0 & \cos{\phi} & -\sin{\phi} & 0 \\
        0 & \sin{\theta}\sin{\phi} & \sin{\theta}\cos{\phi} & \cos{\theta}
    \end{pmatrix},
    \label{eq:spherical_alignment}
\end{equation}
where $\theta = \mathrm{atan2}(z, \sqrt{x^2 + y^2 +z^2})$ and $\phi = \mathrm{atan2}(y,x)$.

Figure~\ref{fig:tetrad_comparison} shows a comparison of each of the three simulation runs, demonstrating that we can accurately model the direction of the beam in each case, although the cylindrical- and spherical-aligned tetrads show some bending due to the rotation of the tetrad in space. This emphasizes the importance of making an appropriate choice of tetrad in order to minimize numerical diffusion and maintain accuracy throughout the simulation.

\subsection{Snake Coordinates \label{sec:snake_test}}

In order to test the accuracy of the angular transport module of our code (corresponding to the third and fourth terms in Equation~\eqref{eq:radiative_transfer_discrete_non_gray}), we make use of snake coordinates, as described in
\citet{White2016}. This coordinate system provides a curved description of space, which serves as an introduction
to more general curved spacetimes. A complete definition from the usual Minkowski coordinates and our choice
of tetrad is provided in Appendix~\ref{app:snake}. For this test, the two parameters of the snake coordinates
are taken to be $A = 0.1$ and $k = 2.0$ corresponding to a maximum variation amplitude of 0.1
and one full period for each unit of $y$.

\begin{figure*}[htbp!]
    \plotone{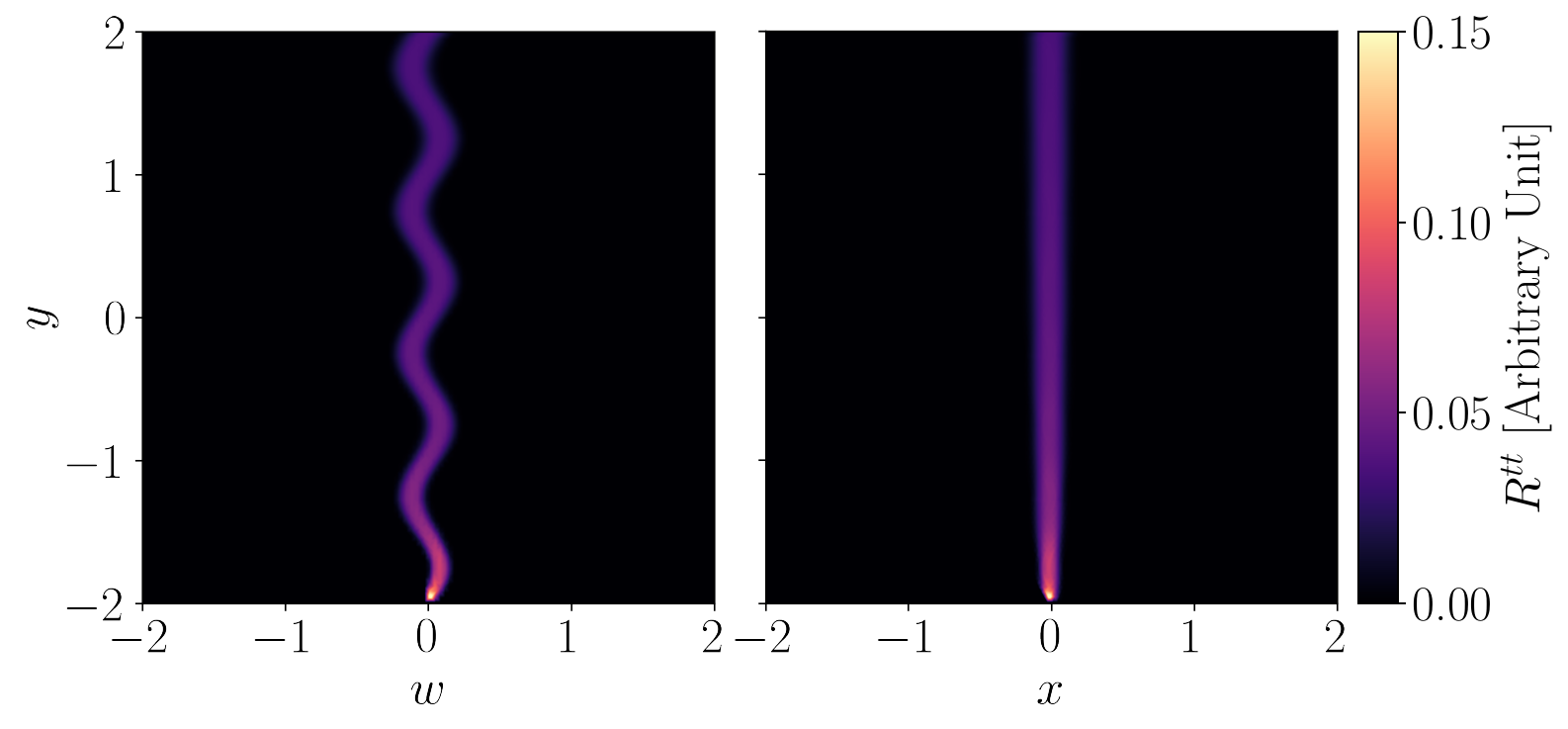}
    \caption{Results of the Snake coordinates test with spatial resolution $128 \times 128$ and G2 angular grid. As expected, the propagation of the light does not appear straight in this coordinate system, as seen in the left-hand panel. The right-hand panel however, which has been transformed back to standard Cartesian coordinates, shows that the beam does indeed follow a straight-line trajectory as expected. \label{fig:snake}}
\end{figure*}

This test was performed with a spatial resolution of $128 \times 128$ and a G2 angular grid. The results are
presented in Figure~\ref{fig:snake}. As anticipated, the propagation of the beam does not appear
as a straight line: the light ray bends, as expected given the rotation of the coordinate system. However, if we
transform back to regular Cartesian coordinates, we can see that the light in this case does indeed travel on a
straight-line trajectory, as would be expected of any light ray.

\subsection{Beam Orbits around Kerr Black Holes \label{sec:circular_orbits}}

Spherical photon orbits around black holes offer an important test of our code, since the ability to resolve
the behavior of a photon beam under relatively constrained conditions demonstrates the numerical
stability and integrity of our method.

In the case of spinning black holes, there are two photon orbit radii, one for co-rotating photons and another for counter-rotating photons. These radii are given by, e.g., \citet{Teo2003} and read:

\begin{equation}
    r_{1,2}\equiv 2\left[1+\cos\left(\frac{2}{3}\arccos\left(\mp a\right)\right)\right]\,,
    \label{eq:photon_orbit}
\end{equation}
where $a$ is the dimensionless spin of the black hole.

This test was carried out by initializing a small strip of cells along these radii, namely in the three cells closest to the circular orbit radius. This results in a beam of radiation with a finite width, where the innermost portions lie inside the circular orbit radius, while the outermost portions are outside this radius. This test therefore verifies that we recover the correct qualitative behavior around this region, where we expect to switch from escaping to infalling radiation as we cross the circular orbit radius.

\begin{figure*}[htbp!]
    \plotone{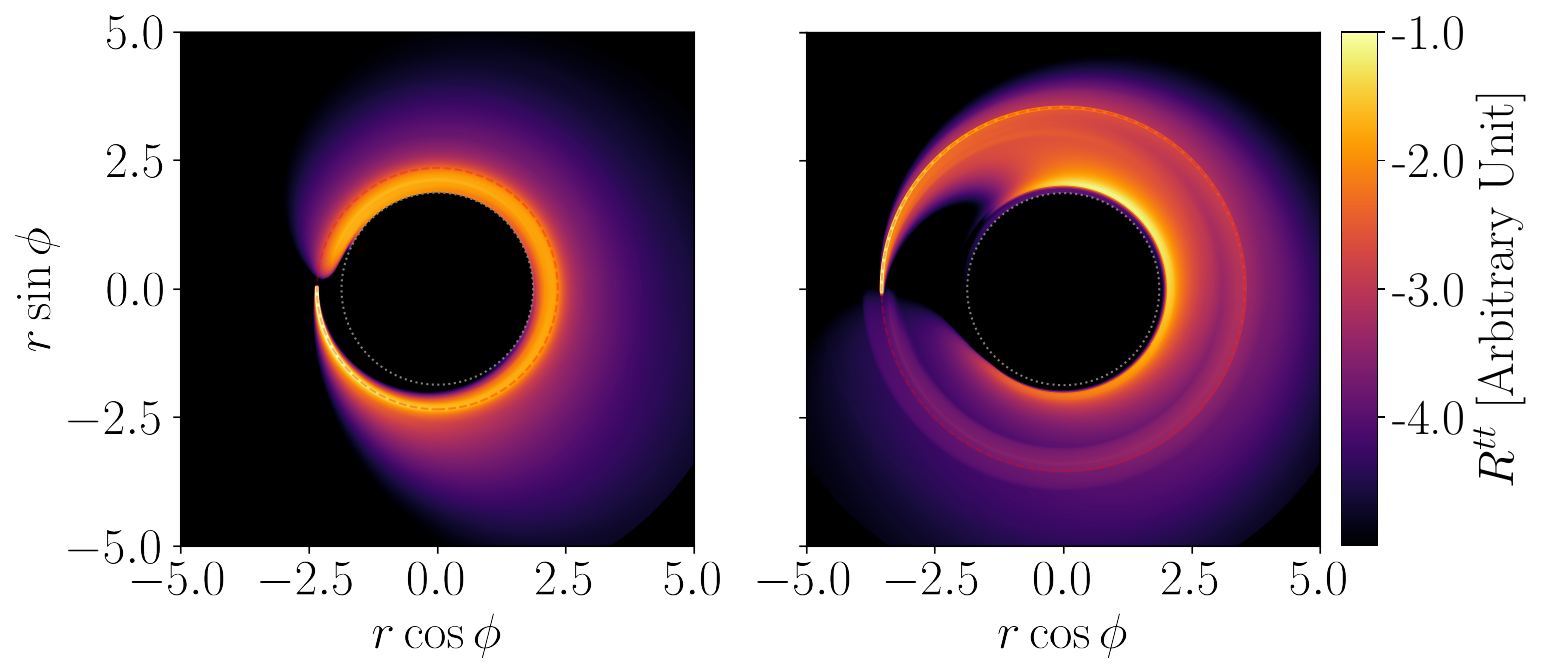}
    \caption{Results of the circular photon orbit test around a BH of spin $a = +0.5$, showing the correct behavior for both prograde orbits (left panel; $r_1 \approx 2.34$) and retrograde orbits (right panel; $r_2 \approx 3.53$). While a circular path is clear, since a strip of cells was initialized around the circular orbit radii given by Equation~\eqref{eq:photon_orbit}, we also show that radiation inside the radius of circular orbits  ($r_{1,2}$) falls inward, while radiation outside the relevant radius escapes outward, thus demonstrating that our radiative transfer scheme correctly handles general relativistic effects close to the black hole.\label{fig:circular_orbits} 
    }
\end{figure*}

We carried out these tests across a number of spatial and angular resolutions, for spins of $\pm 0.5$, $\pm 0.7$ and $\pm 0.95$. In each case, we recover the correct behavior for the beams, with a circular component traveling around the black hole as expected, while the portions of the beam inside the circular orbit radius fall inwards, and those outside the circular orbit escape outwards. In Figure~\ref{fig:circular_orbits}, we show the results of one of these tests, for black hole spin $a=+0.5$.

We use a spatial resolution of 512x256, which we find gives us sufficient resolution in $r$ to properly resolve the radii $r_{1,2}$, and a $\phi$ resolution that is sufficient to show that our scheme properly models the circular path of the trajectory. We use an angular grid of generation 3 (642 angular cells), in which the North pole is aligned with the $\phi$ direction. Since we make use of non-horizon-penetrating coordinates here, namely Boyer-Lindquist coordinates, the increase in $R^{tt}$ seen in the right-hand panel of Figure~\ref{fig:circular_orbits} close to the horizon is an expected feature.


\subsection{Radiating Disk in Vacuum \label{sec:radiating_disk}}

As a final demonstration of the radiation transport scheme, we carry out a test designed to incorporate the full general relativistic capabilities of \cuharm{}.

Our ``disk'' consists of a strip of cells along the equator, which radiates with the temperature expected for the standard Novikov-Thorne model \citep{NT73}:
\begin{multline}
    T = (4 \times 10^7\ \K)\ \alpha^{-1/4} \biggl( \frac{M}{3\Msun}\biggr)^{-1/4} 
    \\ \times r^{-3/8} A^{-1/2} B^{1/2} E^{1/4},
    \label{eq:Novikov_Thorne_temp}
\end{multline}
where:
\begin{subequations}
    \begin{align}
        A & = 1 + \frac{a^2}{r^2} + \frac{2 a^2}{r^3}, \\
        B & = 1 + \biggl(\frac{a^2}{r^3}\biggr)^{1/2}, \\
        E & = 1 + \frac{4 a^2}{r^2} - \frac{4 a^2}{r^3} + \frac{3 a^4}{r^4}.
    \end{align}
\end{subequations}

We use $a = 0.9375$ and $M = 3\Msun$ to parameterize the black hole, and we choose $\alpha = 0.05$ to characterize the disk viscosity. The disk itself extends from $r = 6 r_g$ to $r = 10r_g$. We set the radiation from the disk to be isotropic in the tetrad frame, with:
\begin{equation}
    \Ihat = \frac{1}{4 \pi} \arad \hat{T}^4.
    \label{eq:disk_intensity}
\end{equation}
The radiation is then allowed to evolve in vacuum, (i.e., without interacting). The problem is evolved in horizon-penetrating Kerr-Schild coordinates with the north pole of the geodesic grid aligned with the $\hat \phi$ direction of the tetrad, chosen to be that of the non-rotating observer, see Appendix \ref{sec:takahashi_tetrad} and \citet{Tak08}.

\begin{figure*}[htbp!]
    \plotone{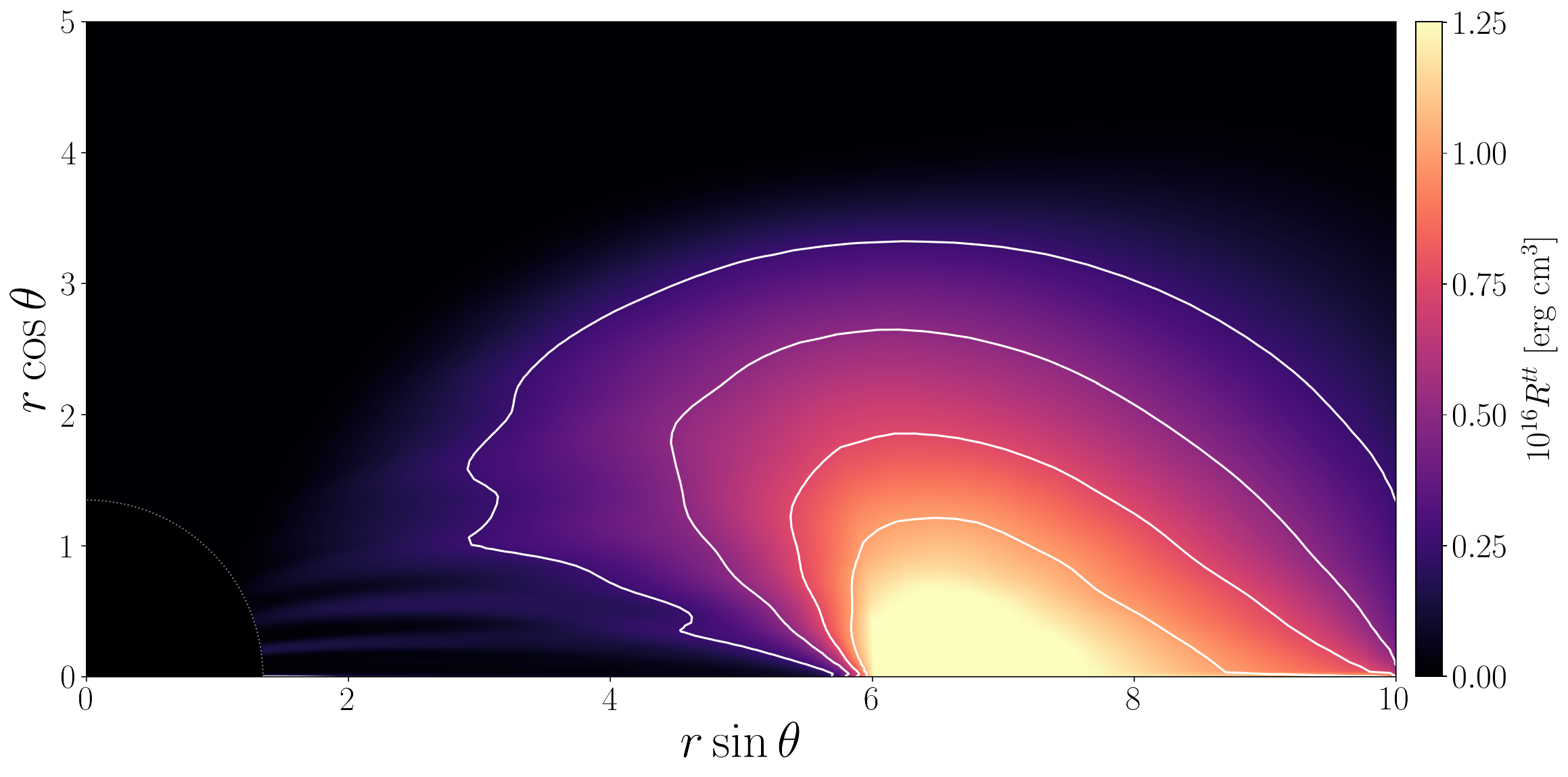}
    \caption{Results of the radiating disk test at $t=5$. Radiation is emitted from the disk midplane (located at $r \cos{\theta} = 0$ on this plot), which extends from $6r_g$ out to $10r_g$. Contours are placed every 0.25 units in $R^{tt}$. In this test, we use a spatial resolution of $128 \times 128$ and a G2 angular grid.\label{fig:radiating_disk} 
    }
\end{figure*}

The results of this test are presented in Figure~\ref{fig:radiating_disk}, which can be compared with those obtained
by \citet{RD20} and \citet{WMJ23}. We note also that there are some ray effects (barely) visible in the numerical solution,
seen in Figure~\ref{fig:radiating_disk} as the thin lines of radiation moving towards the black hole. These rays are
an unavoidable consequence of the angular discretization in this method, although this effect could be minimized by
using a higher angular resolution.

\subsection{Static Radiative Diffusion and Advection}

Up to this point, we have demonstrated that our numerical implementation of radiation transport performs as intended
in vacuum. When considering an MHD setup with a background medium of non-zero density, there exist regions of large
scattering opacity, through which the radiation diffuses. It is important to show that our scheme can recover the
correct diffusive behavior in this limit. Indeed, at high scattering opacity $\alpha^s \gg 1$, the radiative
transfer equation reduces to the diffusion equation for the radiation energy density $E$, here given in 1-dimension,
\begin{align}
    \partial_t E - \frac{1}{3 \alpha^s} \partial^2_x E = 0. \label{eq:diffusion_equation}
\end{align}
The radiation flux is small, on the order of:
\begin{align}
    F^x = -\frac{1}{3 \alpha^s} \partial_x E.
\end{align}

We consider a 1D non-magnetized, cold and non-moving plasma slab from $x = -1$ to
$x = 1$, with constant density, such that $\alpha^s = 10^3$. This hydrodynamical profile is not evolved in this
test and the radiation 4-force is set to zero.

For the radiation, we initialize it such that the radiation energy density is given by 
\begin{align}
    E(x, t = 0) = \exp \left (-\frac{x^2}{2 \sigma^2}  \right ) 
\end{align}
with $\sigma^2 = 1.25 \times 10^{-2}$. Considering this initial
profile, the analytic solution of Equation~\eqref{eq:diffusion_equation} is:
\begin{align}
    E(x,t) = \left ( 1 + Dt \right )^{-\frac{1}{2}} \exp \left ( -\frac{x^2}{2 \left ( 1 + Dt  \right )\sigma^2  }\right ), \label{eq:diff_analytic}
\end{align}
where $D = 2/(3 \alpha^s \sigma^2)$.
In practice, to avoid numerical errors caused by very small values of $E$, we initialize the energy density to be constant for $ |x| > 0.5$, setting it equal to $E(x = 0.5, t = 0)$.

\begin{figure}
\centering
\includegraphics[width=0.45\textwidth]{"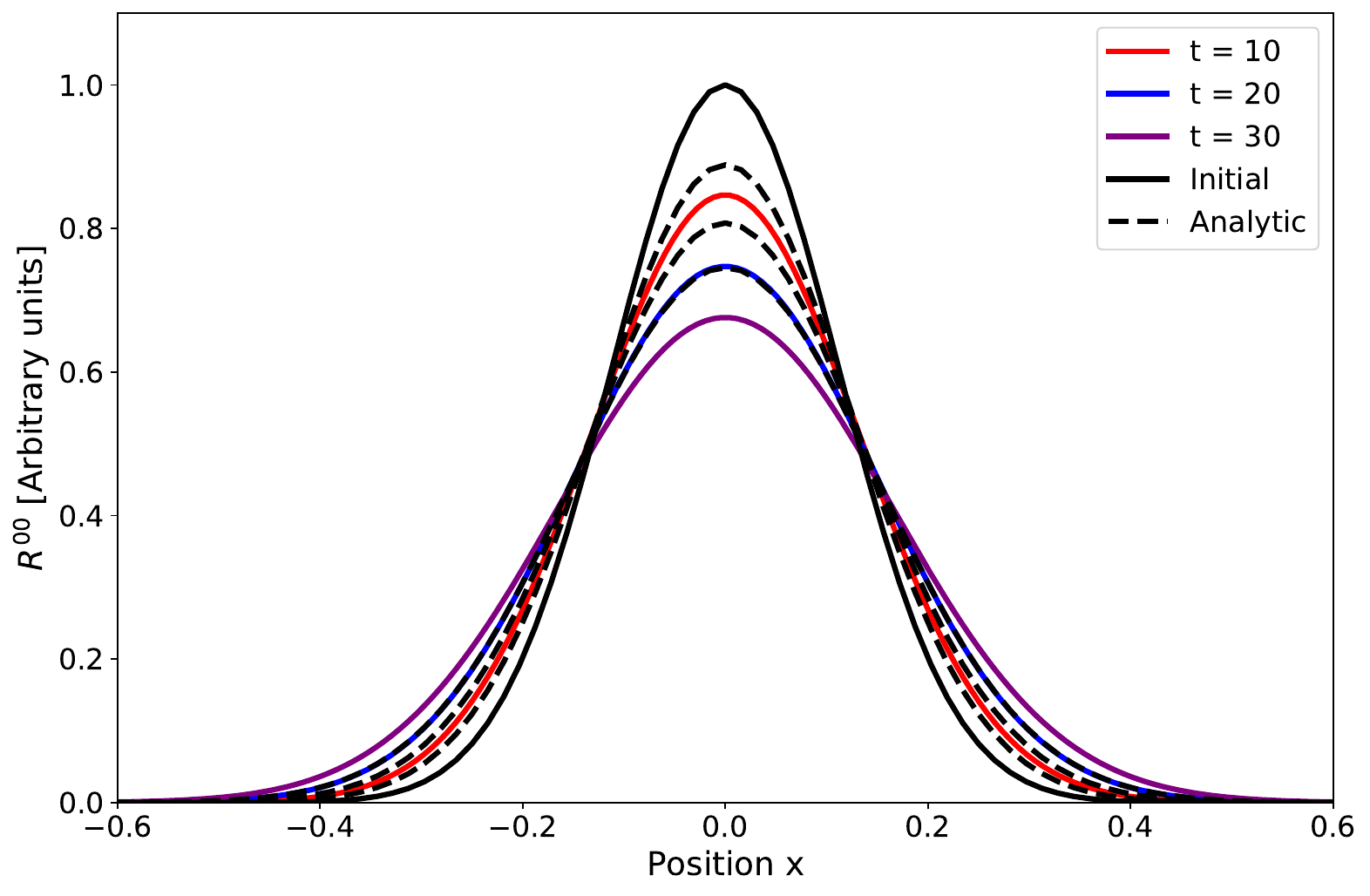"}
\caption{ Static diffusion test with 3 spatial resolutions $N_x \in \{64,128,256\}$ and second generation geodesic grid G2 at times $t = 10,~20$ and 30. The numerical solutions with different resolution are almost indistinguishable. The dashed lines correspond to the analytical solution given by Equation~\eqref{eq:diff_analytic} at the three considered times.}
\label{fig:static_diffusion}
\end{figure}

We run this test with several spatial resolutions, $N_x \in \{ 64,128,256\}$, and G2 for the angular grid generation.
As explained in Section~\ref{sec:fluxes} and appendix~\ref{app:fluxes}, we want to verify that the geodesic grid
does not pose problems for recovering the diffusion limit, and therefore we do not reduce the angular discretization to
only 4 or 8 radiation directions, as in \citet{WMJ23} and  \citet{MPJ25}, respectively.
We show in Figure~\ref{fig:static_diffusion} the result of this test at times $t = 10, 20$ and 30. We find that the numerical
solution depends very weakly on the resolution, and we therefore do not show different resolutions on the figure, as the lines overlap. We do see a systematic offset between the analytic solution and the
numerical one, most likely originating from the choice of $\bar a$ in the definition of the effective opacity (see below Equation~\eqref{eq:rad_speeds}). We did not try to adjust this coefficient to obtain better performances for this test. This
demonstrates that our scheme allows for diffusion to take place and reproduces well the
analytic solution to the 1D diffusion problem, without reducing the number of angles involved in the discretization
of the photon direction. This is an important aspect, since our geodesic grid is not symmetric, as explained in
Section~\ref{sec:fluxes} and Appendix~\ref{app:fluxes}.

Next, we test dynamical diffusion. We consider the same domain and nearly the same initial radiation profile (shifted
to be centered on $x = -0.5$) as for
the previous problem. We further assume that the flow is uniformly and constantly moving with
velocity $v^x = 0.05$. This flow velocity remains constant in the simulation as we turned off the radiation back reaction
for this test as well. Considering the large opacity to scattering, the radiation should be both advected with the
flow while simultaneously diffusing sideways, such that the analytic solution of the problem at time $t$ is given by Equation~\eqref{eq:diff_analytic} with the substitution $x \rightarrow x + 0.5 + v^x t $.

\begin{figure}
\centering
\includegraphics[width=0.45\textwidth]{"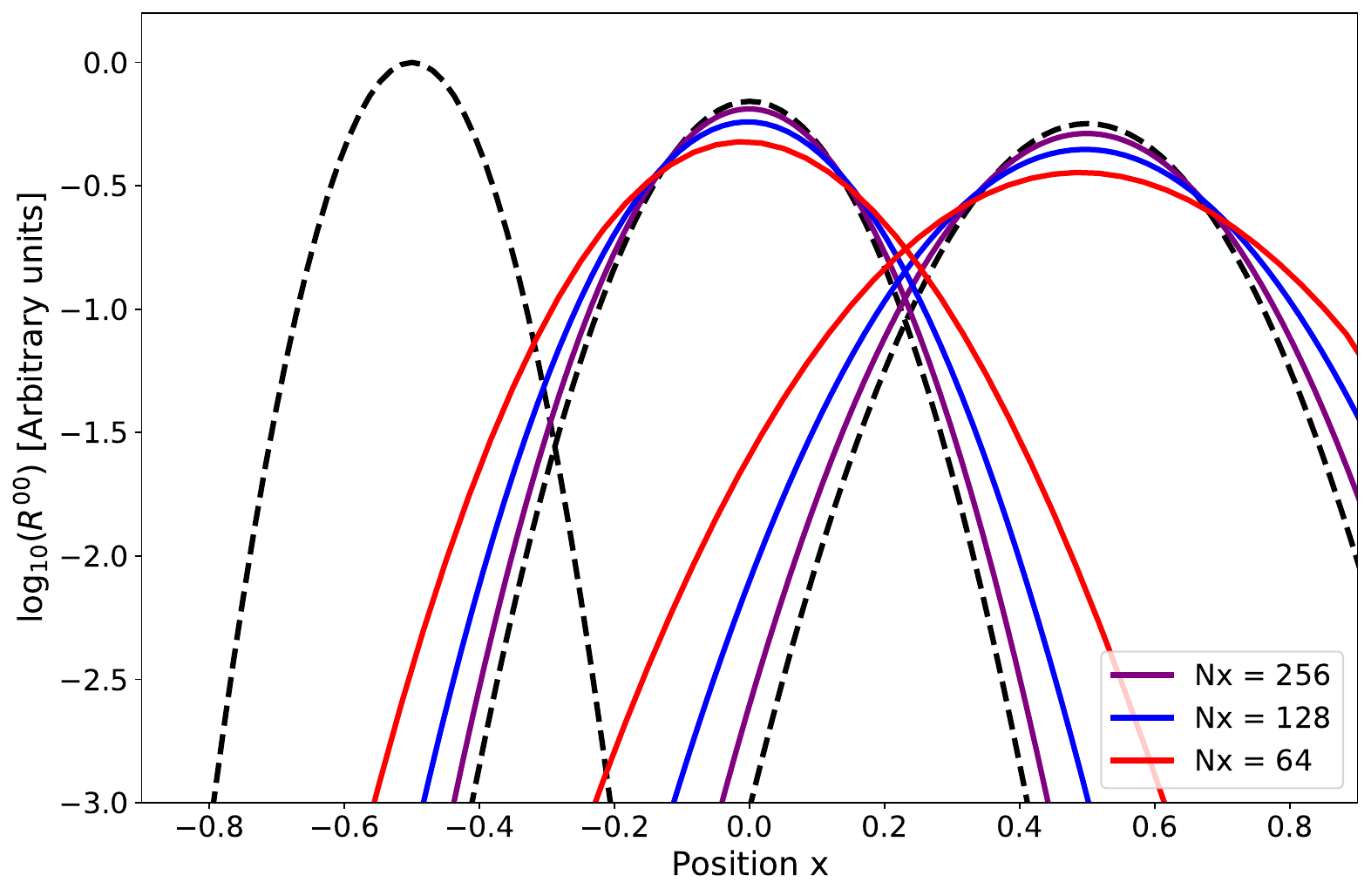"}
\caption{Diffusion and advection test with 3 spatial resolutions $N_x \in \{64,128,256\}$ and second generation geodesic grid G2 at times $t = 10$ and $t = 20$. The three dashed lines correspond to the analytical solution given by Equation~\eqref{eq:diff_analytic} with the substitution $x \rightarrow x + 0.5 + v^x t $, at times $t=0,10,20$. In this test, the numerical solution depends on the resolution via advection.}
\label{fig:advection_diffusion}
\end{figure}

We show the results of this test in Figure \ref{fig:advection_diffusion}, alongside the analytic solution at
times $t = 10$ and $t = 20$. Clearly the radiation profile is advected with the flow at the correct rate since the $x-$positions
of the radiation energy density peaks match that of the analytic solution. Contrary to the diffusion test, the numerical
solution for this test depends on the spatial resolution, due to the advection.

\subsection{Radiation-Matter Equilibration}

Following on from the above collection of tests for the radiation transport scheme, we test the interaction between the radiation and GRMHD sectors of \cuharm.

\begin{figure}[htbp!]
    \plotone{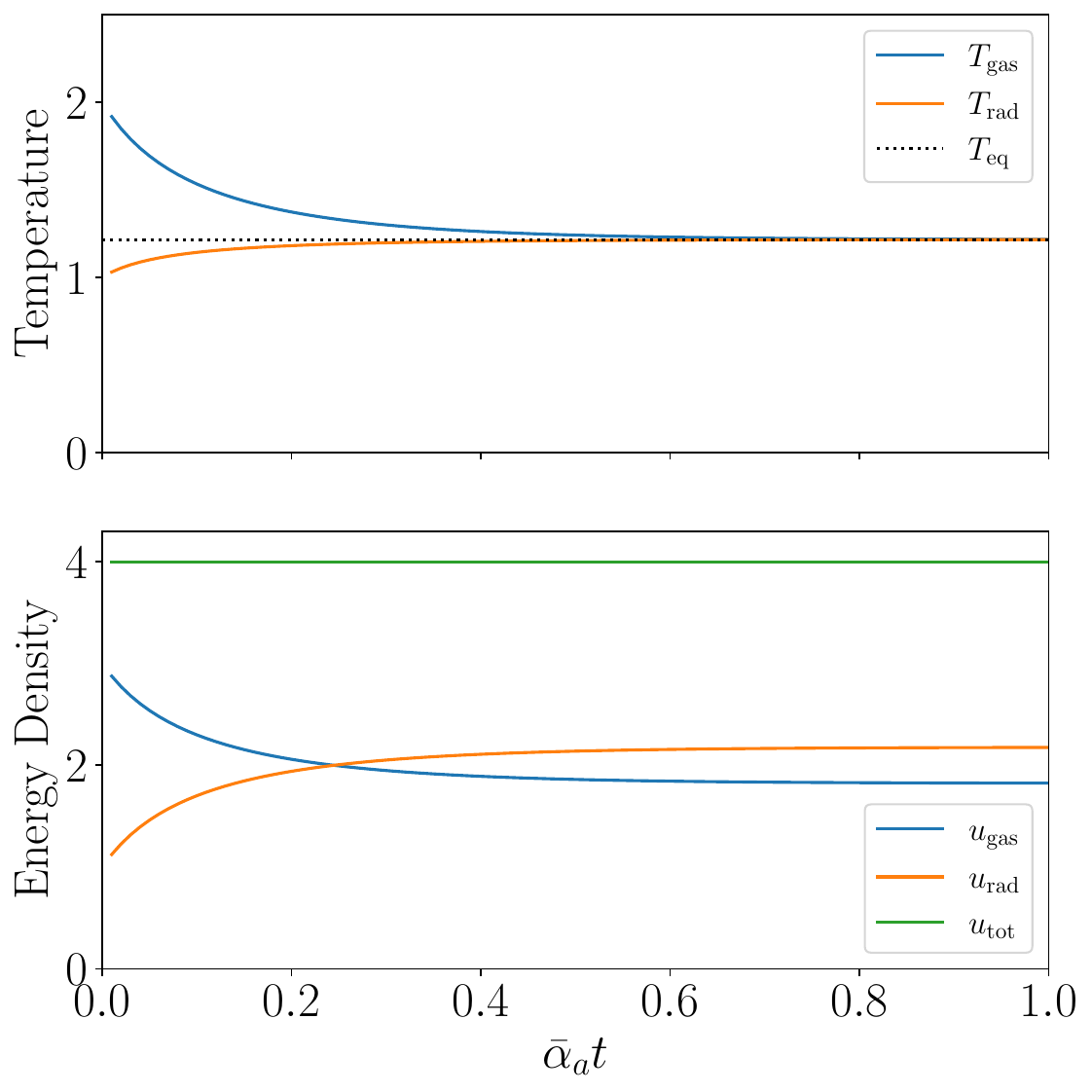}
    \caption{Results of the radiation-matter equilibration test, demonstrating that within a single spatial cell, both radiation and matter settle to the correct equilibrium temperature $T_\mathrm{eq}$, shown here by the black dashed line in the top panel. In the bottom panel, we show that the total energy is conserved within the cell during the equilibration process, where radiation and matter exchange energy until they reach the equilibrium temperature $T_\mathrm{eq}$. \label{fig:equilibration_scale_free}}
\end{figure}

Firstly, we perform a scale-free test of the interaction, namely, we choose units where $G = c = \arad = \kB/{\mu \mprot} = 1$. This allows us to verify the correctness of our interaction algorithm without interference from the natural scaling induced by the black hole mass.

In this system of units, $T_{\mathrm{gas}} = p_{\mathrm{gas}}/\rho = (\Gamma - 1) u_{\mathrm{gas}} / \rho$, and therefore the total internal energy density is:
\begin{equation}
    u = u_{\mathrm{rad}} + u_{\mathrm{gas}} = T_{\mathrm{rad}}^4 + \frac{\rho T_{\mathrm{gas}}}{\Gamma - 1},
\end{equation}
which we use to calculate the equilibrium temperature $T_\mathrm{eq}$.

We consider the temporal evolution of a single localized and isolated cell under the effect of interaction. We set $\Gamma = 5/3$ and initialize the system with $T_{\mathrm{gas}} = 2$, $T_{\mathrm{rad}} = 1$, $\rho$ = 1 and $p_{\mathrm{gas}} = 2$ (which is equivalent to $u_{\mathrm{gas}} = 3$ for our chosen value of $\Gamma$). As the interaction scheme is fully local, we do not employ radiation transport between cells, but instead focus on the temporal evolution of the radiation field in a single spatial cell.

The results of this test, performed with a second generation G2 geodesic grid, are presented in
Figure~\ref{fig:equilibration_scale_free}. The system settles to the correct equilibrium,
and the total energy is conserved throughout the duration of the simulation. Over time, the
temperature of the gas decreases as it emits photons, while the effective radiation temperature increases.
The temperature equilibration takes place on a time scale comparable to $\bar \alpha_a^{-1}$.

\subsection{Radiative Shocks \label{sec:shocks}}

\begin{figure*}[htbp!]
    \centering
    \begin{tabular}{ccc}
    \includegraphics[width=0.35\textwidth]{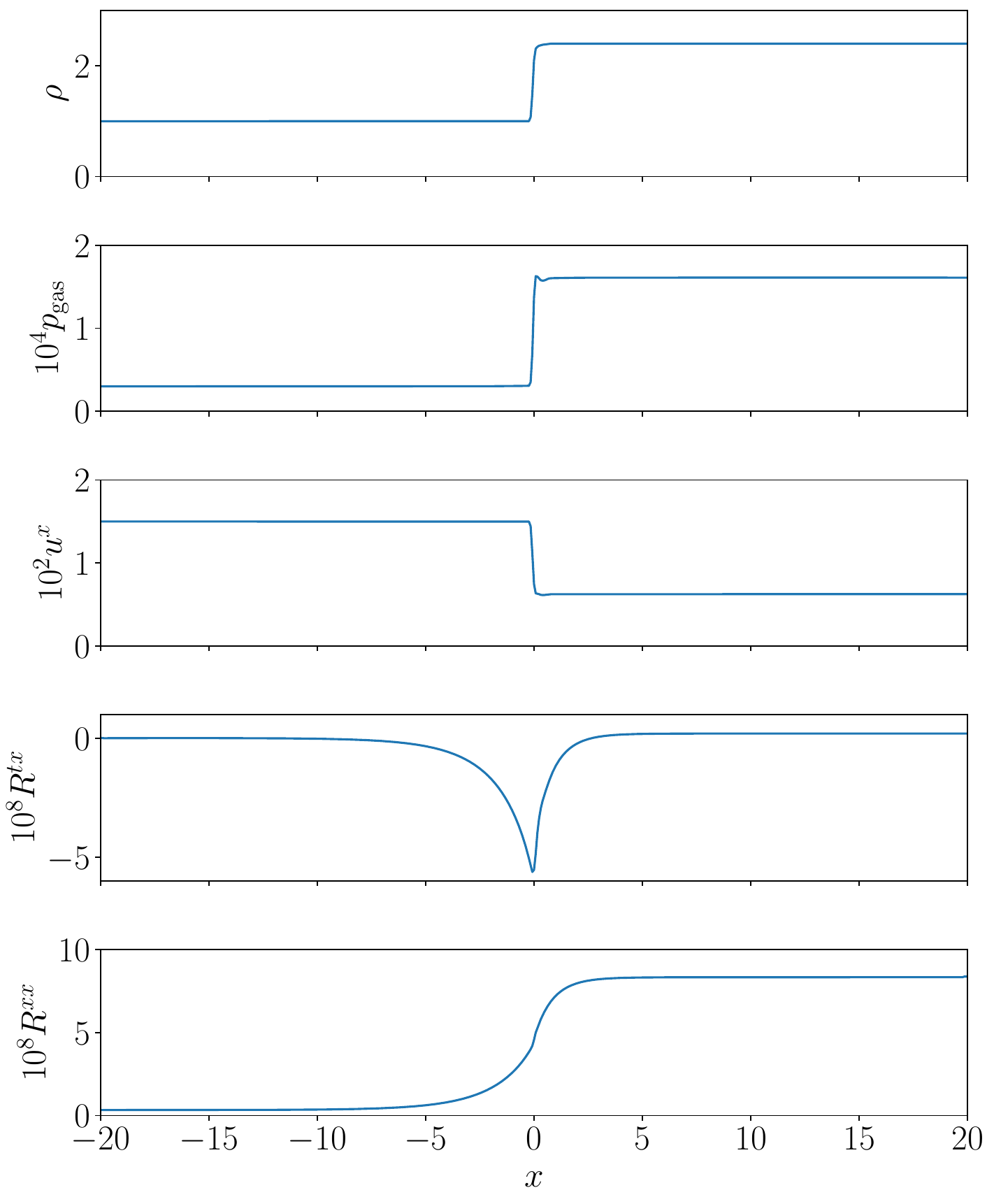}  & ~~~ &
    \includegraphics[width=0.35\textwidth]{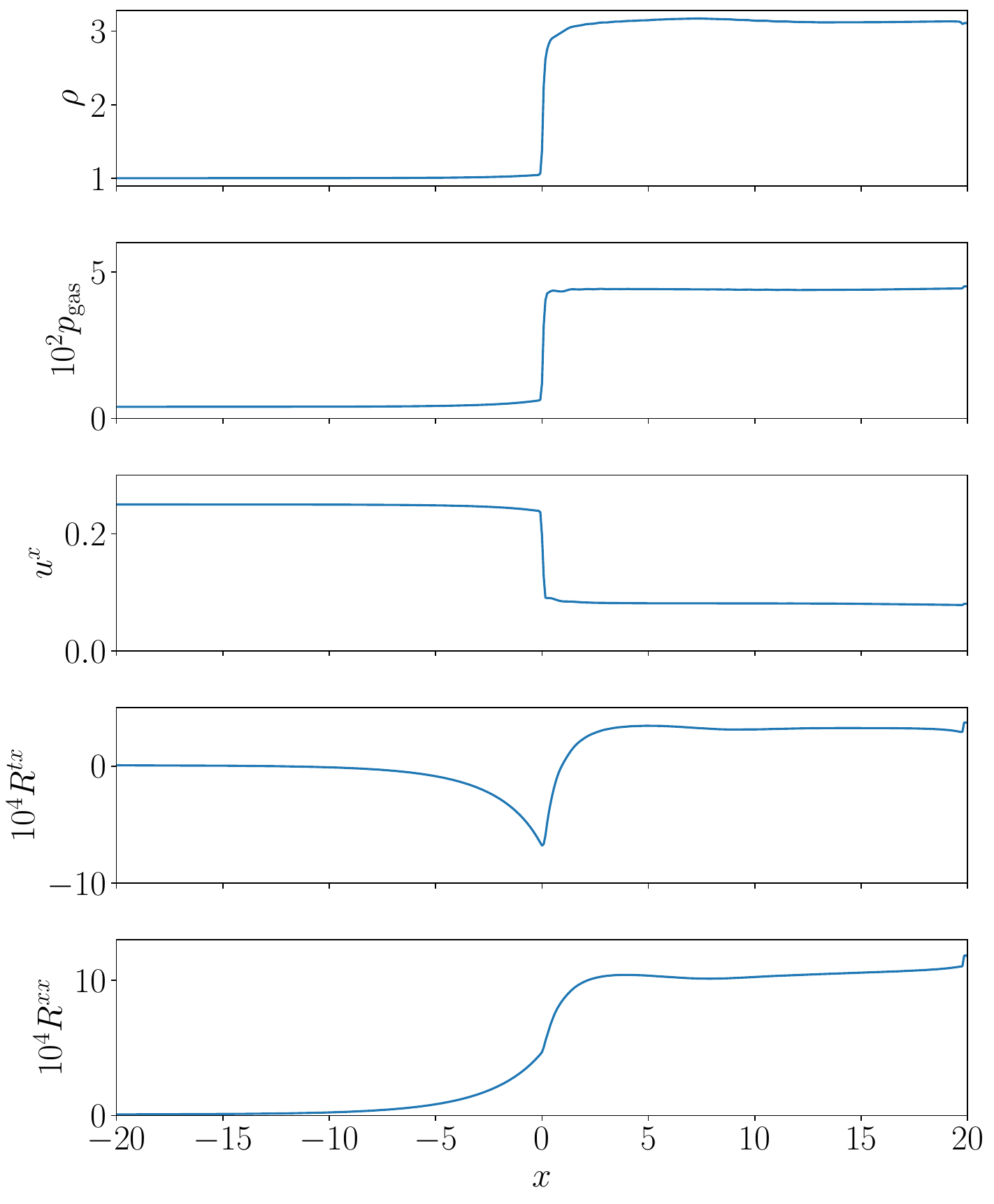}  \\
    ~ & ~ & \\
    \includegraphics[width=0.35\textwidth]{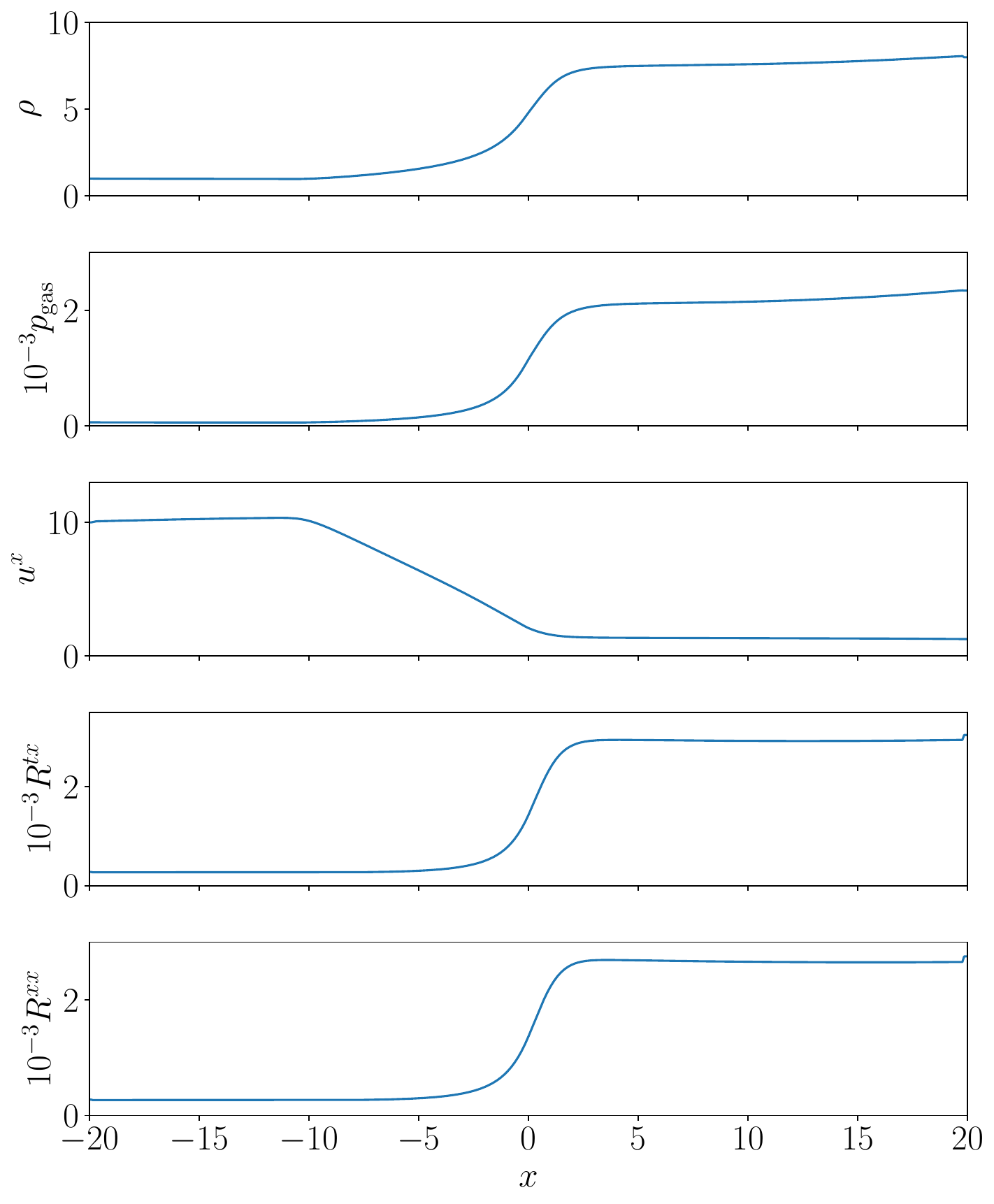}  & ~~~ &
    \includegraphics[width=0.35\textwidth]{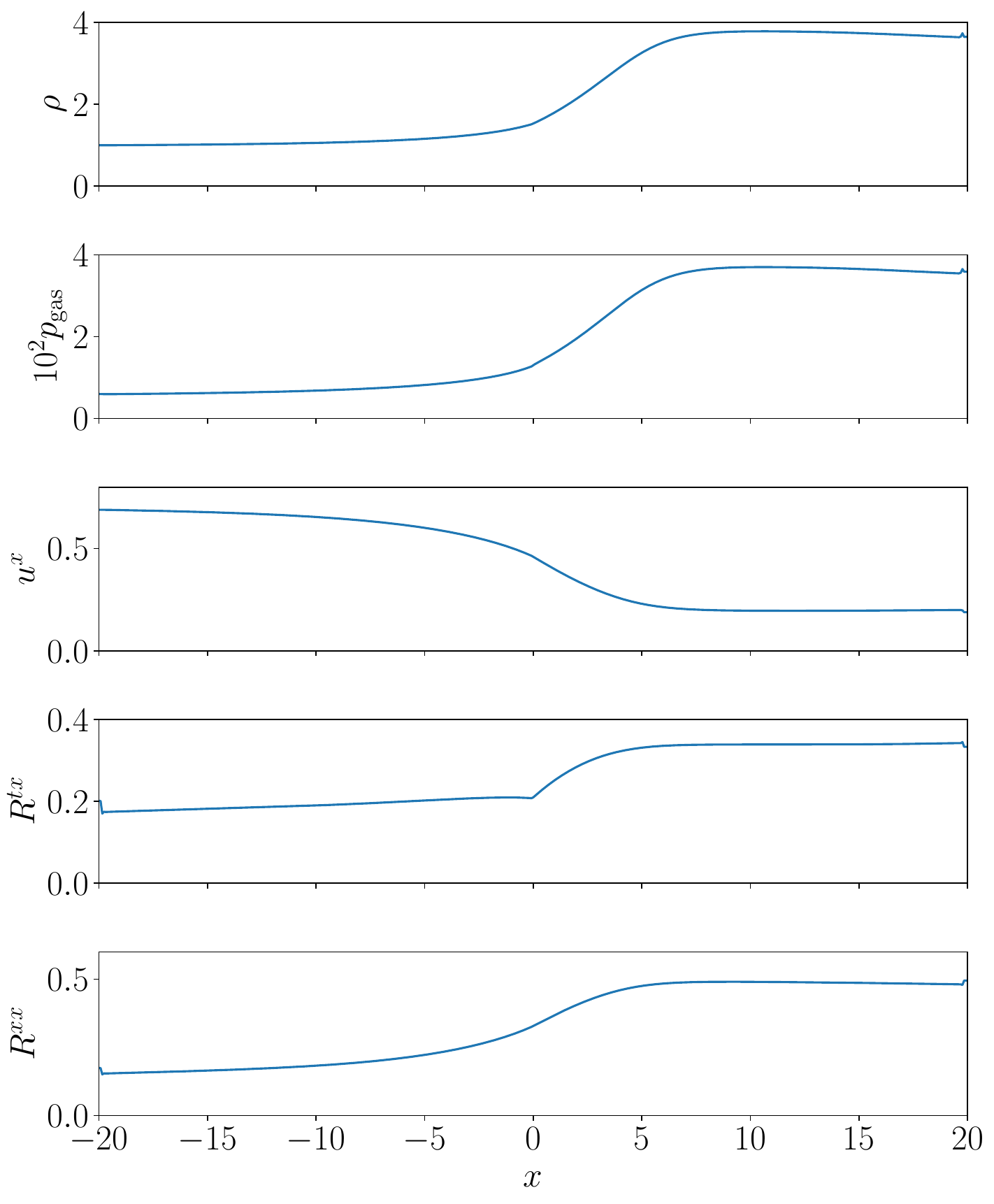}  
    \end{tabular}
    \caption{Results of the 1D radiative shock tests. Top left --- non-relativistic shock test at $t=40$, run with 512 spatial cells and a G2 angular grid. Top right --- mildly-relativistic shock test at $t=400$, run with 512 spatial cells and a G4 angular grid. Bottom left --- highly-relativistic wave test at $t=80$, run with 512 spatial cells and a G4 angular grid. Bottom right --- radiative pressure-dominated wave test at $t=150$, run with 512 spatial cells and a G2 angular grid. For each test, the density, gas pressure, velocity in the $+x$ direction, radiative flux $R^{tx}$ and radiative pressure, $R^{xx}$ are shown.  
    \label{fig:all_shock}
    }
\end{figure*}

As a further test of the interaction between radiation and matter in our code, we perform a simulation of four separate shock scenarios, as first described in \citet{Farris2008}. These scenarios have since become standard tests for radiation-hydrodynamics codes, and consist of: (i) a non-relativistic strong shock, (ii) a mildly-relativistic strong shock, (iii) a highly-relativistic wave and (iv) a radiation pressure-dominated, mildly-relativistic wave.

We set up each scenario as a shock tube, using the parameters given in \citet{Farris2008} (see their Table~1), with the boundaries held at these asymptotic values, following e.g. \citet{Zanotti2011, RD20}. The results of each test are presented in Figure~\ref{fig:all_shock}, whose caption describes the spatial and angular resolution for each tests. In each case we find good agreement with the expected results, which can be compared with, e.g., \citet{RDG15, OT16, RD20, WMJ23}.





\section{Full GR-R-MHD Simulation of an Accretion Torus} 
\label{sec:results}

Having shown that the radiation-matter coupling behaves as expected, we move to a full-scale
test, which is carried out on a realistic setup for a BH accretion disk. Here, we
initialize a Fishbone-Moncrieff torus \citep{FM76} with inner radius $r_{\mathrm{in}} = 6M$ and radius of maximum pressure
$r_{\mathrm{max}} = 12M$. 
Although the system is evolved with radiation, we do not include radiation in
the initial setup, such that initially, the only contribution to the pressure is that of the gas. This is somewhat different from the
approach of \citet{WMJ23}, who assume that initially, the total pressure is due to the contribution of both the gas and radiation
in thermodynamical equilibrium. 
In our approach, the disk gradually relaxes to a thinner accretion configuration as radiation is emitted.

\begin{figure*}[t!]
    \centering
    \begin{tabular}{cc}
    \includegraphics[width=0.45\textwidth]{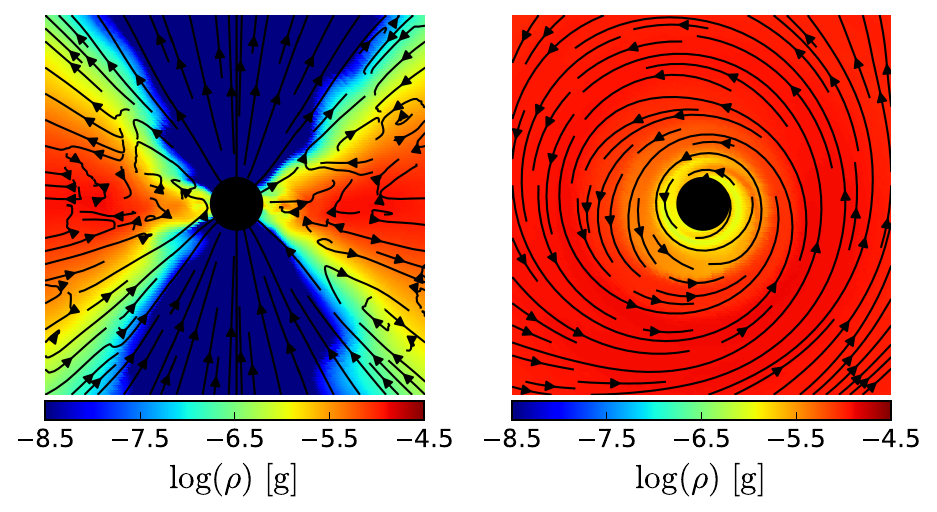} &
    \includegraphics[width=0.45\textwidth]{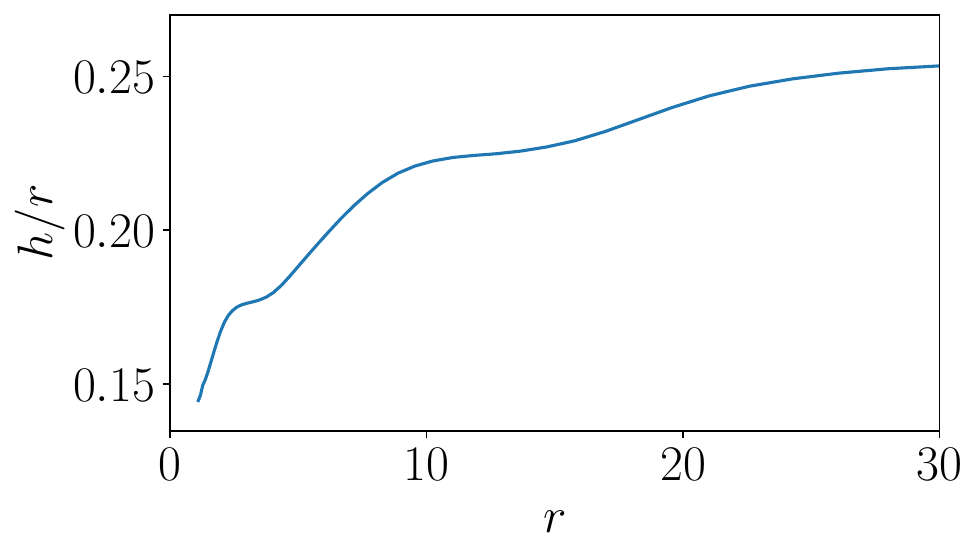} \\
    \includegraphics[width=0.45\textwidth]{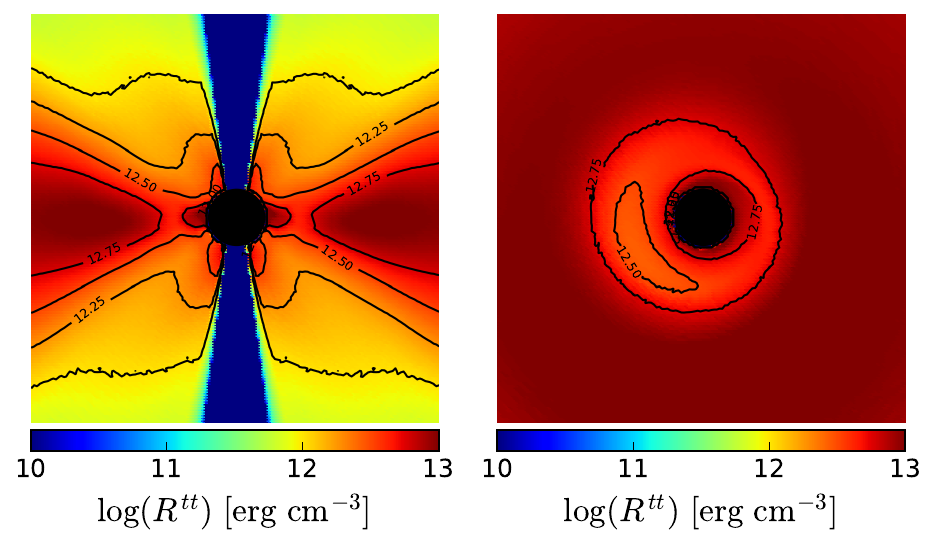} &
    \includegraphics[width=0.45\textwidth]{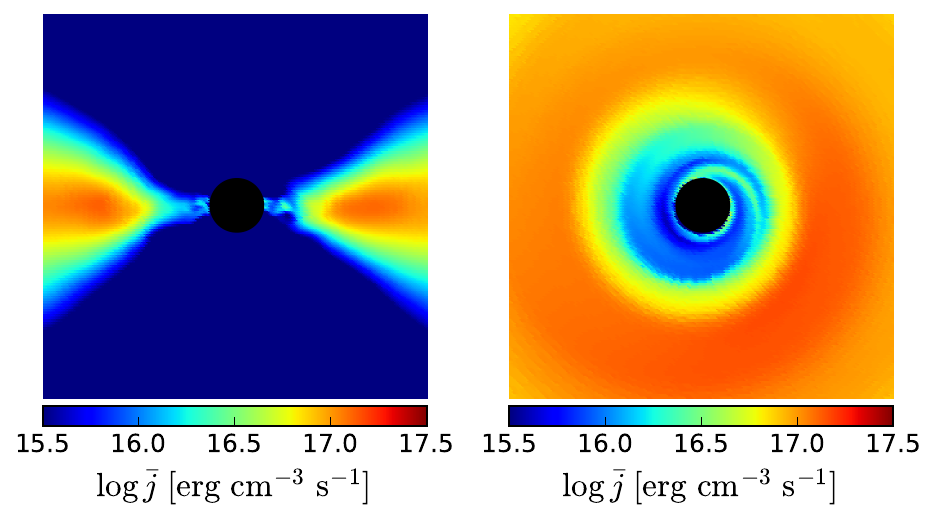} 
    \end{tabular}
    \caption{Results of the torus test problem. Top left --- Polar and equatorial density profile
    for the radiative simulation of the accretion disk around a BH with normalized spin $a = 0.9375$,
    at $t=9250~t_g$. The color coding represents the density. The vector field on the left-hand plot corresponds to
    the magnetic field lines, while on the right-hand plot it represents the velocity. In this simulation, the
    measured mass accretion rate is $1.3 \times 10^{-2} \dot{M}_{\rm Edd}$.
    Top right --- disk height $(h/r)$ as a function of radius $R$, averaged from $t=5000~t_g$ to $t=10000~t_g$, in
    the inner regions of the accretion disk. Bottom left --- Polar and equatorial profile of the coordinate frame radiation energy density $R^{00}$ in erg~cm$^{-3}$, at $t=9250~t_g$. Bottom right --- Polar and equatorial profile of the total fluid frame emissivity $\bar{j}$ (see Equation~\ref{eq:emissivity_total}) in erg~cm$^{-3}$~s$^{-1}$, at $t=9250~t_g$. Videos of the temporal evolution for the mass density and of the radiation energy density can be found on youtube at this \href{https://youtu.be/9JHVpb5sTw8}{url} and this \href{https://youtu.be/xRNyAFOj3gs}{url}, respectively.}
    \label{fig:Torus_density}
\end{figure*}

We use horizon-penetrating Kerr-Schild coordinates to describe the spacetime in the region
around the rotating BH, which we take to have mass $M_{\mathrm{BH}} = 10 \Msun$ and
normalized spin $a = +0.9375$. For the tetrad, we use the tetrad provided by \citet{Tak08}, corresponding
to the locally non-rotating reference frame. The geodesic grid has its north pole aligned with the $x^{\hat 3}$
direction. For the boundaries, we use inflow and outflow conditions in
the radial direction, at the lower and upper boundaries respectively, periodic boundaries in the azimuthal angle, $\phi$, and symmetric
boundaries for the polar angle, $\theta$. We also zero the specific intensity in the four cells closest to the poles. 

For the magnetic field, we initialize a single poloidal magnetic field loop known to lead to a
geometrically thick SANE disk \citep[Standard And Normal Evolution, see e.g.][]{NSP12,SNP13} in the absence
of radiation. In a SANE disk, the magnetic field is always subdominant down to the closest regions of the BH,
but is responsible for providing the viscosity enabling accretion via the magneto-rotational instability
\citep{BH91, BH98}. This process has been numerically demonstrated by means of GRMHD simulations
\citep[see e.g.][]{SMN08,NSP12,CN22}.

The magnetic field is initialized from the vector potential given by
\begin{align}
    A_\phi = \max \left ( \frac{\rho}{\rho_{\rm max}} - 0.2 \right ),
\end{align}
where $\rho_{\rm max}$ is the maximum density on the grid, which is here normalized to unity.
The magnetic field is then normalized such that the ratio of the gas pressure maximum to the magnetic pressure maximum is $\beta_0 = p_{g,  {\rm max}}/p_{b,  {\rm max}} = 100$. As the disk relaxes from its
initial setup to its steady state, radiation is emitted at the expense of the gas energy density, which cools, reaching $T_e \approx 10^{10}$~K in the disk mid-plane. 

In order to set the unit scales according to Equation~\eqref{eq:unit_scale}, we must also set the accretion rate; here, we use
$\dot{M} = 1.4\times10^{18} ~\g ~\mathrm{s}^{-1}$, targeting an accretion rate of 10\% of the Eddington accretion rate for a $10 \Msun$ BH. Normalizing the initial density in code units to
1 and using this expected mass accretion rate, leads to a measured average accretion rate of $1.3 \times 10^{-2} \dot{M}_{\rm Edd}$. We note that the initial hydrodynamical profile does not need to be renormalized; in fact, in \cuharm{}, only the radiation sector uses cgs units, while the evolution of the hydrodynamical sector is calculated in normalized units.

The interaction between fluid and radiation requires the prescription of a temperature model for the electrons. Such a prescription can be
derived  from (i) the primitive MHD quantities, i.e., $\rho, u_g, b^2$. The resulting prescription can be based on either (1) the assumption of constant
electron to proton temperature ratio \citep[see e.g.][]{YCM20,SBP25}; or (2) a prescription in which the electron to proton temperature ratio depends on the plasma-$\beta$ parameter
\citep[see e.g.][]{MFS16}. Alternatively, a prescription can be considered based on (ii) a full 2-temperature evolution
\citep[see e.g.][]{ RRD17, RTQ17, SWN17, LMT22, DMF23, JMF23}. 
For our current 
simulation, we use the approximation of constant electron to proton temperature, such that $T_p / T_e = 1$ everywhere.
This assumption means that the cooling rate is overestimated, as this prescription leads to consistent overestimation of
the electron temperature by assuming full coupling between protons and electrons.

In the simulation we show here, the spatial resolution is $ 128 \times 64 \times 64$, and a second generation (G2)
angular grid, corresponding to 162 angles, is used. In principle, this low polar resolution does not allow to accurately resolve the disk scale height, but is sufficient for the purpose of demonstration and code validation. We run the simulation up to $t= 10^4~t_g$.

\begin{figure}[t!]
  \plotone{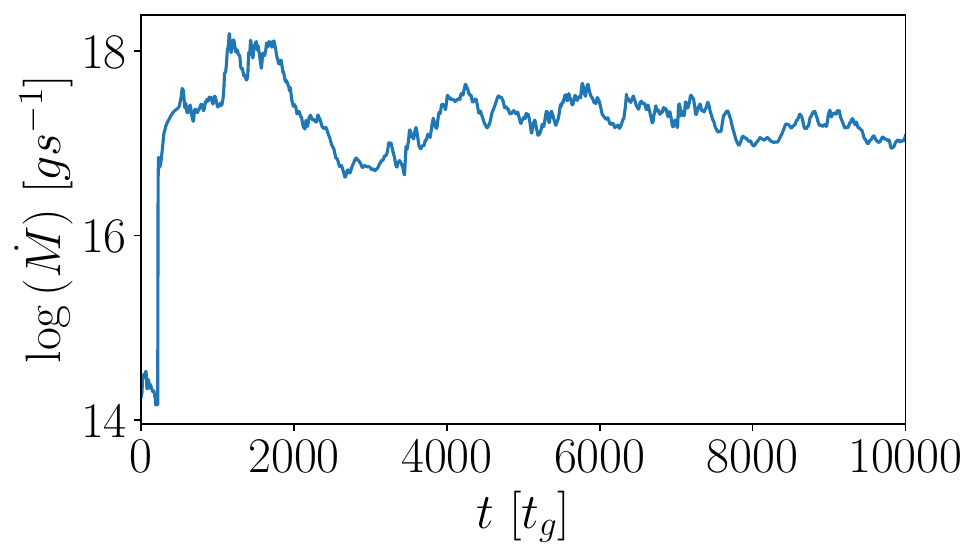}
  \caption{Time evolution of the mass accretion rate in g~s$^{-1}$. After an initial sharp rise corresponding to the onset of accretion, the mass accretion rate oscillates around $10^{17}$~g~s$^{-1}$.  \label{fig:mdot}} 
\end{figure}

The results of our simulation are presented in Figures \ref{fig:Torus_density} and \ref{fig:mdot}.
On the top left panel of Figure~\ref{fig:Torus_density} we show the plasma density for a vertical slice of the domain (containing the
rotation axis of the BH, left) and in the equatorial plane (right) at time $t=9250~t_g$. The vector field represents the magnetic
field lines in the left-hand plot, while on the right-hand plot it represents the velocity. 
The disk is thinner
and denser than its non-radiative counterpart, of which analysis is presented in \citet{BPZ23}. 
The somewhat low measured accretion rate
$1.3 \times 10^{-2} ~\dot M_{\rm Edd}$ explains why the disk is not colder and thinner, as is expected for higher accretion rate \citep[see][]{ZSM25}.

The top right panel of Figure \ref{fig:Torus_density} shows the disk scale height, defined as
\begin{align}
    h/r = \frac{\int \sqrt{-g} \rho \left |\theta - \frac{\pi}{2} \right | \diff \theta \diff \phi }{\int \sqrt{-g} \rho  \diff \theta \diff \phi },
\end{align}
where we note that this ratio is a function of the radius, $r$. 
In the region closest to the BH, the disk is quite thin, before becoming thicker at larger radii. This
is due to the fact that the effective accretion rate at which the disk operates is too low to have triggered the
collapse of the disk. 

Finally, the bottom panels of Figure~\ref{fig:Torus_density} show the energy density $R^{tt}$ in the
radiation field (left) and the emissivity (right). These two quantities are larger in the disk. On the one hand,
the opacity inside the disk is rather large, such that radiation remains trapped in the disk, allowed
to leave only by diffusion. On the other hand, radiation is also mostly produced inside the disk, since outer regions at
the disk boundary (namely the wind) and the jets are not heavily loaded with particles and therefore radiation
production in these regions is small.

We show the time evolution of the mass accretion rate in Figure~\ref{fig:mdot}. After a short transition period, the mass accretion rate is fairly steady with fluctuations by a factor of $\sim 2$.

\section{On the Numerical Intensity}
\label{sec:before_conclusion}

Combining radiation and (magneto-)hydrodynamics via the direct solution of the radiative transfer equation, as done in \cuharm{},  is
numerically challenging. There are several bottlenecks that must be addressed. The basic ones are: (i) large RAM impact; (2) large
required compute throughput; and (3) scaling of the interaction term. Here, we describe these aspects.

\subsection{Radiation as a Memory and Compute Bottleneck.}
\label{sec:rad_bottleneck}

With the inclusion of radiation, the dimensionality of the problem rises from 3 spatial dimensions evolving in time, to 6 (3 spatial and 3 momentum).
Restricting to gray radiation, as presented in this analysis, lowers the dimensionality to 5 (3 spatial and 2 directional),
since the frequency spectrum of the radiation is not considered. Even with this reduction in dimension, the low-resolution
radiative simulation presented in Section~\ref{sec:results} requires about 40~GB of RAM. 

Including the additional
energy dimension will substantially increase the required memory by a factor of at least a few tens. 
This is due to the large dynamical range of the electron temperature, which can vary by 4 to 5 orders of magnitude across the simulation domain. To accurately calculate the resulting spectrum, the photon energy grid
must be resolved over a large range of frequencies, at least twice that of the electron temperature range. This requirement originates from the broad spectra produced by the various radiative processes. Inverse Compton scattering by relativistic electrons ($\theta_{el} \equiv k_B T_e / m_e c^2 > 1$) results in a very broad spectrum, as well as synchrotron emission, which furthermore is typically emitted at lower frequencies, down to the radio band.

The finite volume method allows for a relatively small number of energy bins per frequency decade compared to the finite
difference method. Indeed, the frequency integrals over the interaction kernels can be understood as average interaction
rates over each bin. This is as opposed to the finite difference method which must have a dense enough
frequency sampling to accurately calculate the interaction rate. This advantage was demonstrated in several kinetic studies of
plasma interaction processes, see e.g. \citet{PW05, GBS22}. 

Nonetheless, a minimum of $\sim5$ bins per energy decade
is required to accurately describe the different interactions, implying a memory requirement about 50 times greater for resolving the radiation field over 10 orders of magnitude in
frequency. This would imply a total memory footprint of $\sim$ 2~TB. In order to mitigate this issue, the memory footprint and calculation cost
could be reduced by lowering the angular resolution from G2 to G1, which reduces the memory footprint by a factor of $\sim 4$.

Implementation of full anisotropic and energy dependent Compton scattering is also challenging, although fully
covariant methods have already been designed to perform this calculation in the context of GRMHD \citep{ZW13}. Not only is
the implementation intricate, but combining numerical stability and calculation efficiency is difficult. Indeed, there are two major
aspects to be considered: First, the (potentially) vastly different time scales between the MHD evolution and the
cooling of the electrons requires an implicit solution for the interaction kernel. It is not immediately clear if the method
used in Section~\ref{sec:coupling_calc} can be straightforwardly extended to the full Compton scattering kernel. If this
approach is not immediately applicable, resolving the interaction term fully implicitly would then require solving,
everywhere in space, a non-linear system with a very large number of unknowns, composed of the product between solid angles
on the geodesic grid multiplied by the number of spectral bins, $N_{\Omega} \times N_\nu$, for
the radiation field, and the additional unknown electron temperature. The non-linearity arises due to Compton scattering, whose
kernel is composed of the product of the photon spectrum in two different energy and angular cells. This leads to the
second challenging aspect of such a scheme: the tradeoff between computational efficiency and memory requirement will require
careful treatment, since efficient numerical solution of the non-linear system requires the calculation and storage of
the Jacobian, which has dimension $ \left(N_{\Omega} \times N_\nu\right)^2$, which cannot naively be saved everywhere in space
simultaneously. Alternative Jacobian-free methods may be necessary to overcome this obstacle.

\subsection{GPU Calculation and Implementation}
\label{sec:GPU_implementation}

All calculations presented in this paper are performed on GPUs, which contain massively parallel processors optimized for high-throughput numerical computations. \cuharm{} is written in CUDA-C, which allows it to run on high-end GPUs and benefit
from these powerful capabilities. While naive GPU implementations provide a good
speed-up, dramatically higher performance on GPUs comes at the expense of an important time investment in code
profiling and improvement. This ultimately leads to drastically smaller computation wall-time, through a detailed
balance between efficient memory throughput and compute throughput.

In the current version of \cuharm{} used to produce the results of this paper, although we have naively
improved a few kernels which dominate the calculation time, a major improvement of the computational throughput
remains possible. Currently, we have only implemented (1) the reduction of memory loads by leveraging the fact that the
metric tensor, tetrad vectors and components are independent of the $x^3$ coordinate (corresponding to the azimuthal angle
$\phi$ in Kerr-Schild coordinates), and (2) implementing a reduction for the calculation of the radiation stress-energy
tensor.

From our experience improving the MHD sector, an additional factor of $\sim 10$ is to be expected with (1) a clever
use of shared memory, (2) the gathering of calculations to maximize the ratio of calculations per memory read/store,
and (3) calculation of cells at the boundary of a domain, followed by data transfer between neighboring nodes and GPUs as the calculations in the domain
center are being performed. Since these optimizations are absent in the current version of the radiation module of \cuharm{}, we do not
present code scalings and compute performances, which will instead be provided in a forthcoming publication.



\section{Conclusion \label{sec:discussion}}

We have demonstrated a new radiation module for the \cuharm\ code, which represents a major step forward in the applicability of our numerical model. This radiation sector of the code, in which we solve the time-dependent radiative transfer equation, has been rigorously tested, using several standard tests for a radiation hydrodynamics code, performing well
in each. New regions of parameter space are now accessible for simulation --- in particular, scenarios where radiation
may have a significant effect on the dynamics and evolution of a BH accretion system can
now be accurately modeled without neglecting a major physical component.

In addition to these tests, we have presented the application of our code to a full
general-relativistic BH accretion problem, showcasing the ability of \cuharm{} to track
the evolution of a complex radiation-magnetohydrodynamic system over large dynamical time scales. This
will allow us to perform in-depth study of the effects
of radiation on the structure of BH systems and their evolution. This is particularly timely
in the context of the Event Horizon Telescope (EHT) imagery from both M87 and Sgr A$^*$,
as well as the recently
launched James Webb Space Telescope. Results from both of these observatories will form the basis of the future of
accretion research, and accurate numerical modeling will be central to understanding these observations. It is
therefore of paramount importance to develop accurate, physically complete codes which can be evolved to long
times in a computationally tractable and resource efficient manner. Our GPU-based approach leads to a code which is sufficiently fast to perform the calculation in a reasonable time,
while maintaining numerical accuracy. As detailed in Section~\ref{sec:GPU_implementation}, there remains
room for further optimization and scaling improvement, and the extensible nature of \cuharm\ makes it well-positioned
to make use of the ever-increasing availability of computational resources.

The radiative module of \cuharm{} directly solves the radiative transfer equation on a separate geodesic grid. This offers significant advantages over the commonly used M1 scheme, enabling detailed angular resolution and accurate transport calculation in regions of intermediate opacity. These advantages are crucial in studying systems of complex geometry with extended sources of radiation.

At the present moment, our code can incorporate numerous radiative processes, simply by specifying the relevant
frequency-independent (gray) opacity functions, along with their dependence on the temperature and by further
modifying the formulae for the implicit evolution of temperature to account for these additional temperature dependences, as
described in Section~\ref{sec:coupling_calc} \citep[see also][]{WMJ23, MPJ25}. In the future, we intend to extend \cuharm{} to further include frequency-dependent transport, which adds a significant
computational expense to the current implementation of radiative transfer (see discussion in Section~\ref{sec:rad_bottleneck}), yet provides, for the first time, a complete modeling of BH accreting system and their resulting observational signature.

\begin{acknowledgments}

We acknowledge EuroHPC Joint Undertaking for awarding us access to MeluXina at LuxProvide, Luxembourg.

\end{acknowledgments}






\appendix

\section{Coordinates}
\label{app:coord}
Several coordinate systems have been used in this paper, we provide here their characteristic and our choice of tetrad vectors. 

\subsection{Snake Coordinates}
\label{app:snake}

In terms of the usual Minkowski coordinates $(t, x, y, z)$, snake coordinates can be defined as $(t, w, y, z)$, with $w = x - A\sin{k \pi y}$. Here, $A$ and $k$ are quantities that represent the amplitude and frequency of the coordinate variation respectively.

The metric tensor in snake coordinates is given by:
\begin{equation}
    g_{\delta \gamma} =
    \begin{pmatrix}
        -1 & 0 & 0 & 0 \\
        0 & 1 & \delta & 0 \\
        0 & \delta & 1 + \delta^2 & 0 \\
        0 & 0 & 0 & 1
    \end{pmatrix},
\end{equation}
where $\delta = A k \pi \cos(k \pi y)$. Our chosen tetrad in this case is:
\begin{equation}
    e^{\delta}_{\hat{\alpha}} =
    \begin{pmatrix}
        1 & 0 & 0 & 0 \\
        0 & \delta & 1 & 0 \\
        0 & 0 & 1 & 0 \\
        0 & 0 & 0 & 1
    \end{pmatrix}.
\end{equation}

\subsection{Kerr-Schild Coordinates}
\label{sec:takahashi_tetrad}

For the circular orbit test of Section~\ref{sec:circular_orbits} and the thin disk problem (Section~\ref{sec:radiating_disk}),  we use Kerr-Schild coordinates.
We use the tetrad vectors defined in \citet{Tak08}:
\begin{subequations}
    \begin{align}
        e^{\hat{t}}_{\delta} &= \left[\alpha,~0,~0,~0\right],\\
        e^{\hat{r}}_{\delta} &= \left[\beta^r(\gamma^{rr})^{-1/2},~(\gamma^{rr})^{-1/2},~0,~0\right],\\
        e^{\hat{\theta}}_{\delta} &= \left[0,~0,~(\gamma_{\theta\theta})^{1/2},~0\right],\\
        e^{\hat{\phi}}_{\delta} &= \left[(\beta^r\gamma_{r\phi}\gamma_{\phi\phi})^{-1/2},~\gamma_{r\phi}(\gamma_{\phi\phi})^{-1/2},~0,~(\gamma_{\phi\phi})^{1/2}\right], 
    \end{align}
    \label{eq:coord_to_tetrad}
\end{subequations}
and 
\begin{subequations}
    \begin{align}
        e^\delta_{\hat{t}} &= \left[\alpha^{-1},~-\beta^r\alpha^{-1},~0,~0\right],\\
        e^\delta_{\hat{r}} &= \left[0,~(\gamma^{rr})^{1/2},	~0,~(\gamma^{r\phi} \gamma^{rr})^{-1/2}\right],\\
        e^\delta_{\hat{\theta}} &= \left[0,~0,~(\gamma^{\theta\theta})^{1/2},~0\right],\\
        e^\delta_{\hat{\phi}} &= \left[0,~0,~0,~(\gamma_{\phi\phi})^{-1/2}\right],
    \end{align}
    \label{eq:tetrad_to_coord}
\end{subequations}
where in each case, we define:
\begin{subequations}
    \begin{align}
        \Sigma &= r^2 + a^2 \cos^2\theta \\
        \alpha &= \sqrt{1 + \frac{2mr}{\Sigma}},\\
        \beta^r & = \frac{2mr/\Sigma}{1 + 2mr/\Sigma}\\
        \gamma^{ij} &= g^{ij},\\
        \gamma_{ij} &= g_{ij}.
    \end{align}
\end{subequations}

\section{Details on Numerical Fluxes}
\label{app:fluxes}

In this appendix, we show that the fluxes defined in \citet{Jia21} do not reduce to the diffusion regime when
using a geodesic grid to describe the radiation direction. This prompted us to redefine
the fluxes to accommodate for this regime when using the geodesic grid. For the purpose of this demonstration, we specialize to a
non-relativistic 1D scenario with constant opacity in the domain. The equation of radiative diffusion for the energy density $E$
can be written as
\begin{align}
    \frac{\partial E}{\partial t} = D \frac{\partial^2 E}{\partial x^2 }
\end{align}
where $D = 1/ (3 \kappa_S \rho)$ is the diffusion coefficient. Considering a discretization of this equation on a uniform
grid with grid spacing $\Delta x$, within the framework of the finite volume method leads to the cell-boundary fluxes as
\begin{align}
    F_{i - \frac{1}{2}} = D \frac{E_i - E_{i-1}}{\Delta x} = -\frac{1}{3\tau} \left (E_i - E_{i-1} \right ), \label{eq:flux_left}
\end{align}
where the last equation is obtained by defining the opacity of one cell as $\tau = \kappa_S \rho \Delta x$. The temporal update
then takes the form
\begin{align}
    \frac{\partial E }{\partial t} = F_{i +\frac{1}{2}} - F_{i - \frac{1}{2}} = \frac{1}{3\tau} \left (E_{i+1} + E_{i-1} - 2 E_i  \right ). \label{eq:diffusion_discr}
\end{align}
We will now show that the numerical fluxes proposed by \citet{Jia21} do not reduce to an equation of this form at large opacity.
The fluxes in \citet{Jia21} are similar to those used in our work (see Equation~\eqref{eq:flux_Jiang}), but such that the exponential cutoff in the two first terms
is absent, and the 4$^{th}$ term accounting for advection is not present. In the limit of high opacity, $\tau \gg 1$, the numerical flux
reduces to 
\begin{align}
    F_{i-\frac{1}{2}} = \frac{1}{2}  c \mu_x \left [ I_{i-1}  + I_{i}\right ] - \frac{c |\mu_x|}{2 \tau } \left [ I_i - I_{i-1}\right ],
    \label{eq:flux_Jiang}
\end{align}
where $\mu_x$ is the cosine of the angle between the normal of the face and the direction of propagation. 

The energy density is given by $R^{00}$ and its time variation is
\begin{align}
    \frac{\partial E_i}{\partial t} &= \frac{\partial }{\partial t} \int _{d\Omega} I_i d\Omega = \sum_k \frac{\partial I^k_i }{\partial t} \diff\Omega_k.
\end{align}
We can then use the radiative transfer equation to express the time derivative of $I_\nu$ 
and apply the fluxes in the limit of large opacity to obtain, after rearranging:
\begin{align}
    \frac{\partial E_i}{\partial t} &=  \sum_k \diff\Omega_k \left [ \frac{\mu^k}{2} \left (I_{i+1}^k - I_{i-1}^k \right ) - \frac{|\mu^k|}{2\tau} \left ( I_{i+1}^k - 2 I_i^k + I_{i-1}^k\right )  \right ]. \\
\end{align}
At large opacity, the radiation is nearly isotropic, and therefore we can obtain
\begin{align}
    \frac{\partial E_i}{\partial t}  &= \left (I_{i+1}^k - I_{i-1}^k \right ) \sum_k \frac{\mu^k}{2} \diff\Omega_k   - \left ( I_{i+1}^k - 2 I_i^k + I_{i-1}^k\right ) \sum_k \frac{|\mu^k|}{2\tau} \diff\Omega_k.   \label{eq_app:variation_Ei_no_simplification}
\end{align}
Comparing this expression with the one in Equation~\eqref{eq:diffusion_discr},  it is clear that it is the second
term only which accounts for the diffusion regime, and that the first term should tend to zero or be null when
the radiation field is isotropic. In fact, recalling that
$\int \cos \theta d\Omega = 0$, if the numerical discretization preserves this property, namely
\begin{align}
    \sum_k \frac{\mu^k}{2} \diff\Omega_k = 0 \label{eq_app:int_cos=0}
\end{align}
then we would arrive at
\begin{align}
    \frac{\partial E_i}{\partial t}  \rightarrow   - \left ( I_{i+1}^k - 2 I_i^k + I_{i-1}^k\right ) \sum_k \frac{|\mu^k|}{2\tau} \diff\Omega_k,
\end{align}
where the last sum reduces to $2 \pi$, so that we obtain
\begin{align}
    \frac{\partial E_i}{\partial t}  \rightarrow   - \left ( I_{i+1}^k - 2 I_i^k + I_{i-1}^k\right ) \frac{2 \pi }{2\tau},
\end{align}
which corresponds to the discretization of the diffusion equation up to a constant of the order unity. 

Unfortunately, the geodesic grid used in our work is not symmetric, and as such does not preserve the property given in Equation~\eqref{eq_app:int_cos=0}. In fact, for an angular grid of generation 2, the normalization of the first term of Equation~\eqref{eq_app:variation_Ei_no_simplification} dominates over the second term for an opacity of a few, resulting in failure
of the numerical flux to recover the diffusion regime. We note that this problem does not appear in the diffusion tests performed
by \citet{WMJ23} as their grid is constrained to be symmetrical for these tests.

The numerical flux we propose to use, Equation~\eqref{eq:flux_Jiang}, which is very similar to that of
\citet{Jia21}, does not suffer from this issue, as the first term is exponentially cut off at large opacity, ensuring that the second
term of Equation~\eqref{eq_app:variation_Ei_no_simplification} dominates in the diffusion regime.

\bibliographystyle{aasjournalv7}
\bibliography{cuharm-radiation.bib}



\end{document}